\documentclass[twocolumn]{emulateapj}
\usepackage{mathrsfs}
\usepackage{graphicx} 

\newcommand{\sgr}{Sgr\,A*}

\newcommand{\bldm}[1]{\mbox{\boldmath$#1$}}

\shorttitle{Probing \sgr\ with a pulsars, stars, and the EHT}
\shortauthors{Psaltis et al.}

\begin{document}

\title{A Quantitative Test of the No-Hair Theorem with \sgr\ using
  stars, pulsars, and the Event Horizon Telescope}

\author{Dimitrios~Psaltis$^{1}$, Norbert~Wex$^2$, Michael~Kramer$^{2}$}

\affil{$^{1}$Astronomy Department, University of Arizona, 
             933 North Cherry Avenue, Tucson, AZ 85721, USA}

\affil{$^{2}$Max-Planck-Institut f\"ur Radioastronomie, 
             Auf dem H\"ugel 69, 53121, Bonn, Germany}

\begin{abstract} 
The black hole in the center of the Milky Way, \sgr, has the largest
mass-to-distance ratio among all known black holes in the
Universe. This property makes \sgr\ the optimal target for testing the
gravitational no-hair theorem.  In the near future, major developments
in instrumentation will provide the tools for high-precision studies
of its spacetime via observations of relativistic effects in stellar
orbits, in the timing of pulsars, and in horizon-scale images of its
accretion flow. We explore here the prospect of measuring the
properties of the black-hole spacetime using all these three types of
observations.  We show that the correlated uncertainties in the
measurements of the black-hole spin and quadrupole moment using the
orbits of stars and pulsars are nearly orthogonal to those obtained
from measuring the shape and size of the shadow the black hole casts
on the surrounding emission. Combining these three types of
observations will, therefore, allow us to assess and quantify
systematic biases and uncertainties in each measurement and lead to a
highly accurate, quantitative test of the gravitational no-hair
theorem.
\end{abstract}


\maketitle


\section{Introduction} 
\label{sec:intro}



One of the outstanding challenges in studying theories of gravity is
the experimental verification of the existence of black holes and the
measurement of their fundamental properties.  Based on observational
data, it is generally accepted that at least two types of
astrophysical black holes can be identified, namely stellar-mass black
holes with masses of several $M_\odot$ (see \"Ozel et al.\ 2010 for a
recent compilation) and supermassive black holes with masses between
$10^6$ and $10^{10} M_\odot$ (see Ho 2008 for a review), where
$M_\odot = 2 \times 10^{30}$\,kg is the mass of the Sun.

While supermassive black holes are expected at the centers of most, if
not, all galaxies, in both stellar and supermassive types most of the
observational evidence is based on the interpretation of phenomena
associated with accretion processes onto a
compact object that lacks a hard surface (see, e.g., Narayan \&
McClintock 2013). At the same time, observations of the (gas or
stellar) dynamics around unseen central objects often safely point to
massive objects that are smaller than the Schwarzschild radius
$R_\bullet = 2GM_\bullet/c^2 \sim 3$\,km\,$(M_\bullet/M_\odot)$, where
$G$ is the gravitational constant and $c$ is the speed of light.

Within general relativity (GR), $R_\bullet$ is the equatorial
circumferential radius of the event horizon of an uncharged 
black hole, i.e., the boundary in 
spacetime beyond which
events cannot affect an outside observer. The requirement of an event
horizon surrounding every singularity is trivial to prove for the case
of an uncharged, spherically symmetric spacetime (e.g., the Birkhoff
theorem) but has only been postulated for the most general case in the
{\em Cosmic Censorship Conjecture} \citep{pen79}, which provides the
means to separate the central singularity from the outside world.

In astrophysical situations, black holes are believed to be
(practically) free of any net electrical charge but not without
angular momentum, i.e., black holes are expected to have spin.  In
this case, within GR, the outer spacetime is described by the Kerr
metric\footnote{Strictly speaking, this is only true within a certain
approximation since, to some extent, astrophysical black holes will be
influenced by nearby masses (accreting matter, orbiting objects,
etc.)}, which exhibits an event horizon only for spins less
than a maximum value.  The cosmic censorship conjecture hence requires
for the spin angular momentum $S_\bullet$ of the black hole that
\begin{equation}
  \chi \; \equiv \; \frac{c}{G}\,\frac{S_\bullet}{M_\bullet^2} \; \le 1 \;. 
\label{eq:chi}
\end{equation}

Astrophysical black holes are also expected to be the result of
gravitational collapse of a progenitor and subsequent mass accretion
from the surrounding medium. During the collapse and accretion phases,
all the properties of the incoming material, apart from mass and spin,
are radiated away by gravitational radiation while the gravitational
field approaches exponentially its stationary
configuration \citep{pri72a,pri72b}. Indeed, a powerful uniqueness
theorem within GR has been proven according to which all stationary,
axisymmetric, vacuum spacetimes with no closed time-like loops and no
pathologies outside their horizons are characterized by only three
parameters: the mass ($M_\bullet$), the spin ($S_\bullet$), and the
electric charge (``black holes have no hair''; Israel
1967,1968\nocite{isr67,isr68}; Carter 1971; Hawking
1972\nocite{haw72}; Robinson 1975\nocite{rob75}).

A direct consequence of this no-hair theorem is that all high
multipole moments ($l \ge 2$) of the gravitational field of a
non-charged astrophysical black hole in GR can be expressed as a
function of only $M_\bullet$ and $S_\bullet$ \citep{han74}. In
particular the quadrupole moment, $Q_\bullet$, which is the
lowest-order moment that will be measured observationally, fulfills
the relation~\citep{tho80}
\begin{equation}
  q \;\equiv\; \frac{c^4}{G^2}\,\frac{Q_\bullet}{M_\bullet^3} \;=\; -\chi^2 \;. 
  \label{eq:chiqrel}
\end{equation}
One way of testing the uniqueness of black holes within GR, i.e., the
Kerr hypothesis, and hence the properties of the strongly-curved
spacetime around (spinning) black holes is to measure the mass, spin,
and quadrupole moment of an astrophysical black hole and verify or
refute the above relationship~\citep{rya95}.

The cosmic censorship conjecture and the no-hair theorem
address only a rather limited aspect of strong-field gravity: the
asymptotic, non-dynamical configuration of vacuum gravitational fields
(see discussion in Barausse \& Sotiriou 2008). Verifying them
observationally, however, will increase our confidence in our ability
to use GR in order to predict the outcomes of more general
strong-field gravitational experiments. Perhaps more exciting is the
possibility that either the cosmic censorship conjecture or the
no-hair theorem may be proven not to be satisfied for astrophysical
black holes. Even though violating either or both can be accommodated,
in principle, within GR, such an observational result will most likely
have very serious consequences for the foundations of the theory. This
is especially true since many minimal modifications of the gravity
theory leave the no-hair theorem and the Kerr metric unaffected (see
Psaltis et al.\ 2008; Sotiriou \& Faraoni 2012).

The black hole in the center of our Galaxy provides the optimal
setting for testing the cosmic conjecture hypothesis and the no-hair
theorem with multiple, independent experimental probes (see Psaltis \&
Johannsen 2011 for an early discussion and Ghasemi-Nodehi et al.\
2015).  Optical/IR imaging of the stars in the central region revealed
closed orbits around a central black hole with a mass of about
$4.3\times 10^6
M_\odot$ \citep{2008ApJ...689.1044G,2009ApJ...692.1075G}.  At a
distance of about 8.3\,kpc, the implied apparent size of the shadow
cast by the black hole on the surrounding emission is in the realm of
Very Long Baseline Interferometry (VLBI) observations at
mm-wavelengths (see Falcke et al.\ 2000).

{An international effort is underway to conduct such a mm VLBI
experiment with the Event Horizon Telescope (EHT) that will allow us
to image the shadow of this supermassive black hole, known as
Sagitarius~A* (\sgr), against the background of emission from a hot
accretion disk (Doeleman et al.\ 2009b). Initial EHT observations with
only a minimal set of interferometric baselines have indeed confirmed
the presence of horizon-scale structures in its emission, and
simulations of the full array indicate that true imaging of
strong-field general relatistic signatures will soon be possible (Fish
et al 2014).}  Measuring the shape and size of the black-hole shadow
can be used to infer the mass, quadrupole moment, and (to a lesser
extent) the spin of the black hole and, hence, to test the cosmic
conjecture hypothesis and the no-hair theorem (see, e.g., Bambi \&
Freese 2009; Johannsen \& Psaltis 2010b; Broderick et al.\ 2014;
Psaltis et al.\ 2015b).

The shape of the black-hole shadow is determined by purely
gravitational effects and modeling it does not depend on our
understanding of the accretion flow properties. {Even though
astrophysical effects, such as the presence of opaque plasma in front
of the black hole, might obscure partially the shadow, they will not
affect its shape or size. As such, a test of the no-hair theorem with
the EHT is largely immune to the usual complexities that are
inherent in most astrophysical observations. However, there remains
the possibility of systematic biases in such a measurement caused,
e.g., by the misidentification of the outline of the black-hole shadow
or by an erroneous subtraction of the blurring effects of
interstellar scattering (Psaltis et al.\ 2015b; see also Fish et al.\
2014; Lu et al.\ 2014).}

The good news is that, in the near future, the spacetime of \sgr\ will
be studied in more than one way, at a range of distances and with
different probes by tracing the orbits of stars and pulsars with next
generation instruments. For the former, the adaptive-optics assisted
optical interferometer GRAVITY will have the ability to observe
relativistic effects in the orbits of stars that reach within a few
hundred gravitational radii of the central black hole (Eisenhauer et
al.\ 2011).  The power of the latter is derived from pulsar timing
observations where, even for pulsars with relatively poor timing
accuracy, the instantaneous time-of-arrival (TOA) for a pulsar signal
can be measured with an uncertainty of a few hundred microseconds,
corresponding to a light-travel (``ranging'') distance of only $\sim
10-100$\,km. A phase-connected solution with an appropriate timing
model leads to a determination of the pulsar orbit, which is
considerably better than that. Hence, as argued
by \cite{wk99}, \cite{2004ApJ...615..253P} and \cite{lwk+12}, a pulsar
in orbit around the super-massive black hole in the Galactic Center
would be a sensitive probe to the black-hole properties.  {A
number of international projects are contributing to the EHT effort,
the ERC-funded project BlackHoleCam among them, to exploit the synergy
between probing the properties of \sgr\ using EHT imaging, stellar
orbits, and pulsar timing.}

The motivation of the work presented here is to demonstrate the power
of combining the constraints derived from the EHT observations of the
black-hole shadow to those from independent measurements based on
stellar and pulsar orbits in quantifying and assessing potential
systematic effects in the test of the no-hair theorem.  In the
following we describe the methods for the various experiments and
their prospects, and demonstrate the synergies of the different
approaches. After a general description of the Galactic Center black
hole and its environment for stellar and pulsar orbits, we look
at the possibility of probing the black-hole spacetime with stellar
orbits. We then review and expand on the treatment of pulsar orbits,
before we show how to combine the results of the previous sections
with potential information from the imaging observations of Sgr A*.


\section{General Definitions and Considerations}
\label{sec:gen}



\subsection{The Central Black Hole}

Optical/IR observations of the orbits of stars in the vicinity of
\sgr\ have led to a measurement of its mass, $M_\bullet$, and distance
from the Earth, $D$. The uncertainties in the two measurements are
significant and highly correlated (Ghez et al.\ 2008; Gillessen et
al.\ 2009). Because of the directions of these correlations, however,
the uncertainty in the apparent size of the black-hole shadow, which
is the most relevant quantity for the EHT observations, is
significantly smaller. In the following discussion, we set the mass of
\sgr\ to $M_\bullet=4.3\times 10^6\;M_\odot$ and its distance to
$D=8.3$~kpc \citep{2014ApJ...783..130R}, such that the apparent
opening angle of one gravitational radius ($GM_\bullet/c^2$) at the
distance of \sgr\ is equal to $5.1~\mu$as and consistent with
the most likely value derived from current observations (Psaltis et
al.\ 2015b).

Because of the large orbital distances of the currently known
optical/IR stars around \sgr, there have been no dynamical
measurements of its spin magnitude, $\chi$, or orientation. Comparison
of accretion flow models with spectroscopic and EHT imaging
observations indicate low spins, when semi-analytic models are used
(e.g., Broderick et al.\ 2011), or relatively high spins when GRMHD
models are used (e.g., Dexter et al.\ 2010; Chan et
al.\ 2015). Moreover, the small inferred size of the 1.3\,mm image of
\sgr\ supports the assumption that the black-hole spin is inclined by
$\simeq 50-60^\circ$ with respect to the line of sight and is aligned
with the angular momentum vector of the stellar disk at $\sim
3$~arcsec away from the black hole (Psaltis et al.\ 2015a).  
{For the purposes of the present paper, we set the spin of \sgr\ 
to $\chi=0.6$, which corresponds to a Kerr quadrupole moment of
$q=-0.36$.  We picked these values such that the effects of both the
spin and of the quadrupole moment are potentially observable,
without being maximal.  Clearly, we can perform tests of the no-hair
theorem only if the black hole in the center of the Milky Way is
spinning.}


\subsection{The Inner Cluster of Stellar-Mass Objects}

Advances in adaptive optics have revealed a large number of stars in
orbit around \sgr\ \citep[see][]{2010RvMP...82.3121G,2012mgm..conf..420G}.  
One of these stars has been followed for at least one fully closed orbit
\citep{2008ApJ...689.1044G,2009ApJ...707L.114G} and the orbital parameters
of several others (S0-16, S0-102, and S0-104) will eventually place
them within a few thousand gravitational radii from the black hole
(e.g., Meyer et al.\ 2012). Even though monitoring of these orbits in
the near future will most likely lead to detection of periapsis
precession, additional relativistic effects that will allow for a test
of the no hair theorem will either be too small to be detected or masked
by other astrophysical complexities.

It is expected that observations with future instruments, such as the
adaptive-optics assisted interferometer GRAVITY on the VLT (Eisenhauer
et al.\ 2011) and new generation adaptive optics instruments on a 30-m
class telescope (Weinberg et al.\ 2005), will lead to the discovery of
stars with closer orbits. Monitoring the precession of their orbits and
of their orbital planes will offer the possibility of measuring the spin
and the quadrupole moment of the black hole and, therefore, of testing the
no-hair theorem (Will 2008).

The distribution of stellar-mass objects within a few thousand
gravitational radii from \sgr\ is very difficult to infer
observationally at this point (see the detailed discussion in Merritt
2010). For the purposes of the current study, we will follow Merritt
et al.\ (2010) and set the distribution of the semi-major orbital axes
of stellar objects around the black hole such that
\begin{equation}
  n(a)=\frac{N_0}{a_0^3} \left(\frac{a}{a_0}\right)^{-\gamma}\;.
\end{equation}
{We will write our expressions in the general case of $\gamma<3$, 
but evaluate them in the corresponding figures for $\gamma=2$ (Merritt et 
al.\ 2010) and $\gamma=7/4$ (Bahcall \& Wolf 1976), to quantify the effect 
of this assumed parameter}. Requiring that the total mass of stars inside the
characteristic orbital separation $a_0$ is equal to $M_*$, i.e.,
\begin{equation}
  m_*\int_0^{a_0}4\pi a^2 n(a)da=M_*\;,
\end{equation}
we obtain for the normalization constant
\begin{equation}
  N_0 = \frac{3-\gamma}{4\pi}\frac{M_*}{m_*}
\end{equation}
and for the total number of stars inside an orbit with semi-major axis $a$
\begin{equation}
  N(a) = \left(\frac{a}{a_0}\right)^{3-\gamma}\frac{M_*}{m_*}\;.
\label{eq:Na}
\end{equation}
The characteristic values for the mass $m_*$ of each object and the
total mass $M_*$ enclosed inside an orbital separation $a_0$ are also
poorly constrained from current observations. We will adopt here a
conservative set of values (Merritt et al.\ 2010) for which 
$m_*=1\,M_\odot$, $a_0=1$\,pc, and $M_*=10^6\,M_\odot$.

We can use this distribution to calculate the mass, angular momentum,
and quadrupole moment due to the stellar cluster that is enclosed
inside an orbit of a given semi-major axis. The ratio of these
quantities to the black-hole mass, angular momentum, and quadrupole
moment will represent the limiting accuracies to which these
black-hole properties can be inferred using observations of orbits of
stars and pulsars.

{The mass of stars inside a circular orbit with semi-major axis $a$
\begin{equation}
M_*(a)=m_* \int_0^a 4\pi a^2 n(a)da
\end{equation}
and the relative contribution to the mass of the black hole is
\begin{eqnarray}
\frac{M_*(a)}{M_\bullet}&=&\left(\frac{M_*}{M_\bullet}\right)
\left(\frac{a}{a_0}\right)^{3-\gamma}\nonumber\\
&=&4.8\times 10^{-8}
\left(\frac{M_*}{10^6\,M_\odot}\right)
\left(\frac{a_0}{1\,\mbox{pc}}\right)^{-1}
\left(\frac{ac^2}{GM_\bullet}\right)\,
\label{eq:ClusterMass}
\end{eqnarray}
where in the last expression we set $\gamma=2$.}

The enclosed angular momentum due to the stellar cluster depends on
the relative orientation of the orbits and the distribution of their
eccentricities. We can obtain an upper limit to the enclosed angular
momentum by assuming that all orbits are circular and aligned.  In
this case, the enclosed angular momentum is
\begin{equation}
 J_*(a) = m_* \int_0^a 4\pi a^2 n(a) \left(G M_\bullet a\right)^{1/2}da\;.
\end{equation}
The dimensional spin angular momentum of the black hole is
$S_\bullet\equiv \chi \, G M_\bullet^2/c$ 
(cf.\ equation~[\ref{eq:chi}])
and, hence, the magnitude of the relative contribution to the angular 
momentum due to the stellar cluster is
\begin{eqnarray}
  \chi\frac{J_*(a)}{S_\bullet}&=&\frac{2(3-\gamma)}{7-2\gamma}
  \left(\frac{a}{a_0}\right)^{3-\gamma}
  \left(\frac{M_*}{M_\bullet}\right)
  \left(\frac{ac^2}{GM_\bullet}\right)^{1/2}\nonumber\\
&=& 3\times 10^{-8}
  \left(\frac{M_*}{10^6\,M_\odot}\right)
  \left(\frac{a_0}{1\,\mbox{pc}}\right)^{-1}
  \left(\frac{ac^2}{GM_\bullet}\right)^{3/2}\;,\nonumber\\&&
\label{eq:ClusterSpin}
\end{eqnarray}
{where in the last expression we set $\gamma=2$.}

The enclosed quadrupole moment due to the stellar cluster also depends
on the orientation of the orbits. If we add an axisymmetric angular
dependence to the distribution of orbits, i.e., denote this
distribution by $n(a,\theta,\phi)= n(a)n_\theta(\theta)$, then the
quadrupole mass moment of the stellar cluster becomes
\begin{eqnarray}
  Q_*(a) &=& \frac{m_*}{2}  \int_0^a a^2 n(a,\theta,\phi)
             \left(3\cos^2\theta-1\right) a^2 da\,d\phi\,d\cos\theta
         \nonumber\\
         &=& \frac{\tilde{n}_\theta}{12}\left(\frac{a}{a_0}\right)M_* a^2\;,
\end{eqnarray}
where we have defined
\begin{equation}
\tilde{n}_\theta\equiv \int_{-1}^1
  \left(3\cos^2\theta-1\right) n_\theta(\theta)\,d\cos\theta\;.
\end{equation}
The dimensional quadrupole angular momentum of the black hole is
$Q_\bullet=q\,G^2 M_\bullet^3/c^4$ 
(cf.\ equation~[\ref{eq:chiqrel}]) and, hence, the magnitude of the
relative contribution to the quadrupole moment due to the
stellar cluster is
\begin{eqnarray}
  q\frac{Q_*(a)}{Q_\bullet}&=&\frac{(3-\gamma)\tilde{n}_\theta}{4(5-\gamma)}
\left(\frac{M_*}{M_\bullet}\right)
\left(\frac{a}{a_0}\right)^{3-\gamma}
\left(\frac{ac^2}{GM_\bullet}\right)^2\nonumber\\
&=&4.0 \times 10^{-10}
\left(\frac{\tilde{n}_\theta}{0.1}\right)
\left(\frac{M_*}{10^6\,M_\odot}\right)
\left(\frac{a_0}{1\,\mbox{pc}}\right)^{-1}\nonumber\\
&&\qquad\qquad\qquad\qquad\qquad\qquad
\left(\frac{ac^2}{GM_\bullet}\right)^3\;,
\label{eq:ClusterQuad}
\end{eqnarray}
where in the last expression we set $\gamma=2$.

\begin{figure}[t]
\begin{center}
  \includegraphics[height=7cm,width=8cm]{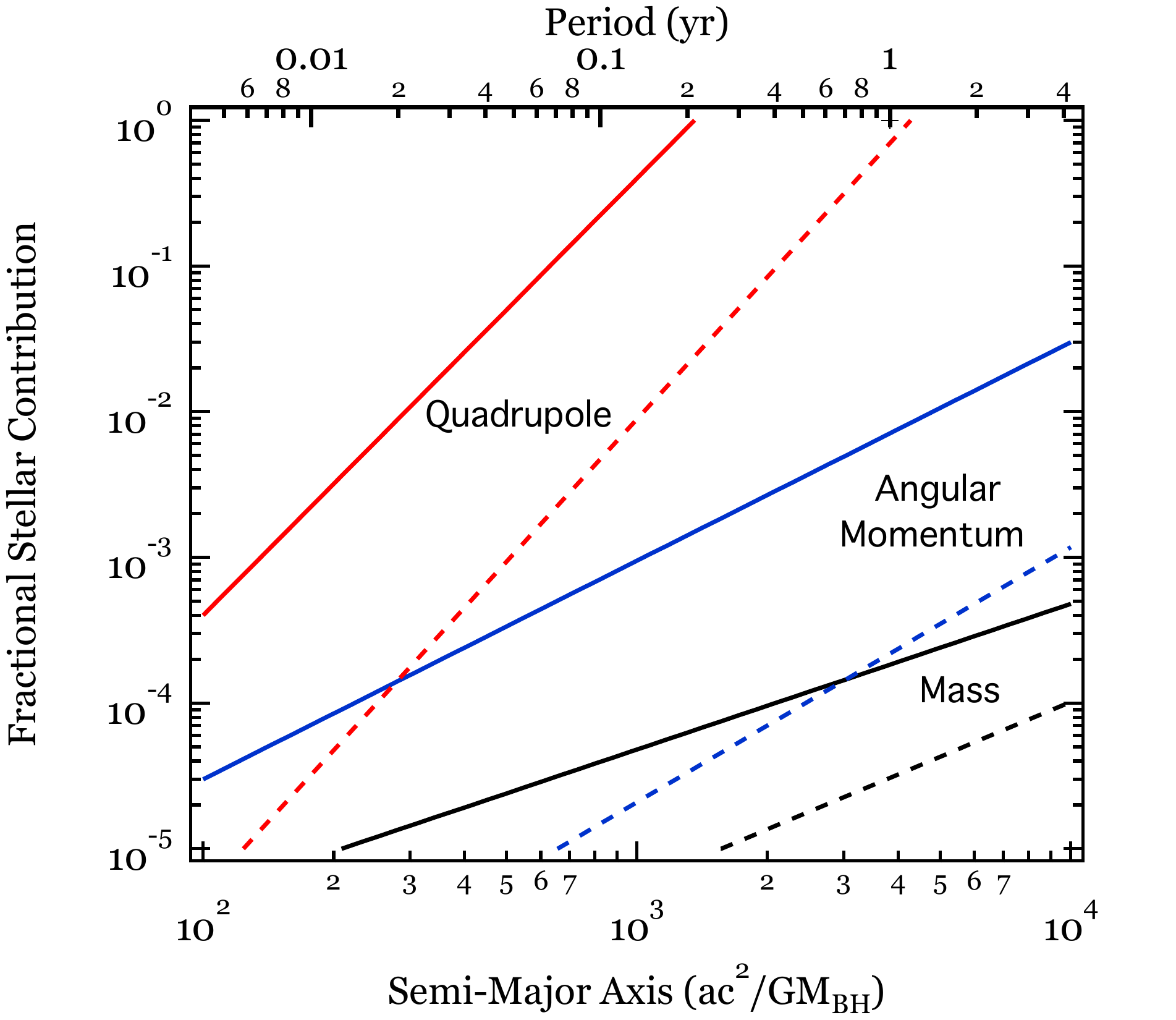}
\caption{Fractional contribution to the black-hole mass, angular
  momentum, and quadrupole mass moment inside an orbit due to the
  enclosed distribution of objects. These fractional contributions
  represent the limiting accuracies to which the corresponding
  black-hole properties can be inferred using observations of orbits
  of stars and pulsars. {The solid lines correspond to a stellar
  distribution with $\gamma=2$, while the dashed lines correspond to
  $\gamma=7/4$.} The various other assumed parameters of the stellar
  cluster are given in equations~(\ref{eq:ClusterMass}),
  (\ref{eq:ClusterSpin}), and (\ref{eq:ClusterQuad}.)
\label{fig:cluster}}
\end{center}
\end{figure}

The fractional contributions to the mass, angular momentum, and
quadrupole moment enclosed inside an orbit of semi-major axis $a$ are
shown in Figure~\ref{fig:cluster}. Our goal is to use orbits of stars
and pulsars to measure the quadrupole moment of the black hole and
test the no-hair theorem. Just imposing the requirement that the
stellar cluster does not dominate the quadrupole moment of the
gravitational field forces us to use circular orbits with orbital
separations (or equivalently elliptical orbits with periapsis
distances) that are inside a few times $\simeq 1000\,GM_\bullet/c^2$
(see also Merritt et al.\ 2010). For pulsars in highly eccentric
orbits ($e \gtrsim 0.8$), as we will demonstrate in
Section~\ref{sec:psr}, we have, besides the secular precession of the
orbit, an additional probe of the relativistic effects via the
near-periapsis periodic contributions, which are less affected by
external perturbations.

\subsection{Pulsars in the Galactic Center}
\label{sec:GCpsrs}

For a number of observational and theoretical considerations, we
expect a large number of neutron stars in the central part of the
Galaxy. For a comprehensive review of the observational evidence and
related theoretical considerations, we refer to \cite{wcc+12} and
references therein. Based on evidence for, e.g., the past star
formation rate, the expected initial stellar mass function in the
Galactic Center environment, and the observations of massive stars and
stellar remnants, overall up to 100 normal pulsars and 1000
millisecond pulsars (MSPs) should be expected in the inner parsec. Earlier,
\cite{fl11} pointed out that the high stellar density in the region
allows also the effective creation of exotic binaries, like
MSP-stellar black-hole binaries, which would be
exciting laboratories in their own right \citep{wk99,lew+14}.

Millisecond pulsars are old, recycled pulsars, which show typical
periods between 1.4 and 30\,ms, while normal pulsars have average
periods of 0.5 to 1\,s. millisecond pulsars also have spin-down rates
and estimated magnetic field strengths that are typically three orders
of magnitude smaller than those of normal (unrecycled) pulsars. These
properties make millisecond pulsars superior -- and hence preferred --
clocks in pulsar timing experiments. For a normal pulsar, a typical
timing precision is around 100\,$\mu$s, while for the best millisecond
pulsars one can achieve a timing precision as good as 100\,ns or
better. In both cases, the final timing precision depends on the
pulsar itself (e.g., the sharpness of its pulse shape, the intrinsic
rotational stability) and the strength of the pulsar, as the error on
an individual time-of-arrival (TOA) measurement scales with the
signal-to-noise ratio of the observation (see Lorimer \& Kramer
2004\nocite{lk04} for further details on pulsar properties and timing
methods).

Despite concentrated efforts and dedicated searches in the Galactic
Center region, the yield has been disappointingly low given the
estimates. Until 2013, only five pulsars had been found within $15'$ of
\sgr, with the closest of these $11'$ away, i.e., at a projected
distance of about 25\,pc (Johnston et al.\ 2006; Deneva et al.\ 2009;
Bates et al. 2011)\nocite{jkl+06,dcl09,bjl+11}. All of these were slow
pulsars with dispersion measures up to 1500\,pc\,cm$^{-3}$. Given their
distances to \sgr, none of these are suitable for the experiments
described below.

The resulting perceived paucity of Galactic Center pulsars had been
explained as a consequence of hyper-strong scattering of the radio
waves at the turbulent inhomogeneous interstellar plasma in the
region. The scattering leads to temporal broadening of the pulses with
expected timescales of at least $2000 (\nu/1$\,GHz$)^{-4}$\,s
\citep{cl02}, rendering their detection impossible at typical search
frequencies, around 1 to 2~GHz. For this reason, a number of
high-frequency searches were conducted in the past (Kramer et
al.\ 2000; Klein et al.\ 2004, Johnston et al.\ 2006; Deneva et al 2010; Macquart et
al.\ 2010; Bates et al.\ 2011; Eatough et al.\ 2013; Siemion et
al.\ 2013) \nocite{kkl+00,jkl+06,mkfr10,bjl+11,ekk+13,sbb+13} at
frequencies as high as 26\,GHz. However, even in these searches, no
pulsar in the central parsec was found. The currently best limit
($S_{min}<10\,\mu$Jy for a $S/N \sim 10$) is provided by observations
with the 100-m Effelsberg telescope at 19\,GHz (Eatough et al., in
prep.).

The recent discovery of radio emission from the magnetar
SGR~J1745$-$29 by Eatough et al.\ (2013; see also Shannon \& Johnston
2013)\nocite{efk+13,sj13}, which had been first identified at X-rays
(Kennea et al.\ 2013; Mori et al.\ 2013), \nocite{kbk+13,mgz+13}
provides an unexpected probe of the Galactic Center medium and the
local pulsar population. The source that, with improved positional
precision, is now named PSR~J1745$-$2900, is located within $2.4''$
(or 0.1\,pc projected) of \sgr\ \citep{bdd+15} and is strong enough 
that even single
pulses can be detected from a frequency of a few GHz \citep{sle+14} up
to an unprecedented 154\,GHz (Torne et al.~2015)\nocite{tek+15}. Below
1.1\,GHz, the temporal broadening prevents a detection of the source
(Spitler et al.~\ 2014), while pulsed radio emission is detected up to
225 GHz, which is the highest frequency at which radio emission from a
neutron star has been detected so far (Torne et al.\ 2015). The
dispersion measure and the rotation measure of PSR~J1745$-$2900 are
the largest in the Galaxy (only the rotation measure of \sgr\ itself
is larger; Eatough et al.\ 2013; Shannon \& Johnston 2013), while the
angular broadening of the source is consistent with that of
\sgr\ \citep{bdd+14, bdd+15}, providing evidence for the proximity of
the magnetar to the Galactic Center. While its rotational stability is
unfortunately not sufficiently good to conduct precision timing
experiments, it allows us to revisit the question of the hidden pulsar
population.

Radio emitting magnetars are a very rare type of neutron stars and
previously only three of them were known to exist in the Galaxy, i.e.,
less than 0.2\% of all radio-loud neutron stars \citep{ok14}. The
discovery of such a rare object adjacent to \sgr\ thereby supports the
notion that many more ordinary radio pulsars should be present
\citep{efk+13,cl14}.  A surprising aspect of the magnetar discovery is
the relatively small scatter broadening that is observed (Spitler et
al.\ 2014). With a pulse period of 3.75~s, its radio emission should
not be detectable at frequencies as low as 1.1~GHz, if hyper-strong
scattering were indeed present.

Imaging observations (Bower et al.\ 2015) resulted in the measurement
of a proper motion that does not allow us yet to conclude as to
whether the pulsar is bound to \sgr. It is possible that
PSR~J1745$-$2900 and the other five nearby pulsars originated from a
stellar disk (see also Johnston et al.\ 2006) and that a central
population of pulsars is still hidden. {Indeed, Chennamangalam \&
Lorimer (2014)\nocite{cl14} argue that, even if the
lower-than-expected scattering in the direction of PSR~J1745$-$2900 is
representative of the entire inner parsec, the potentially observable
population of pulsars in the inner parsec has still a conservative
upper limit of $\sim 200$ members. They conclude that it is premature
to assume that the number of pulsars in this region is small. 

In contrast, Dexter \& O'Leary (2014)\nocite{do14} come to 
 a different conclusion. They also revisited the
  question about the central pulsar population given the new
  constraints provided by the magnetar and the non-detection of
  previous high-frequency surveys.  Considering various effects like
  depletion of the pulsar population due to kick velocities exceeding
  the central escape velocity, pulsar spectra and the apparent reduced
  scattering indicated by the magnetar observations (Spitler et
  al.~2014), they argue in favour of a ``missing pulsar
  problem''. They also concluded that the magnetar discovery in the
  center may imply, in turn, an efficient birth process for magnetars
  in the central region. Similarly,  others suggested that normal
  pulsars are not formed since they may collapse into black holes on
  comparably short timescales by accreting of dark matter
  \citep{bl14}. 

  At the core of deciding between these possibilities is our ability
  to properly model and account for all selection effects in the
  previous surveys. There are in fact indications that this is not the
  case. Firstly, continued monitoring of the scattering timescales for
  the magnetar indicates that the scattering time is highly
  variable. While it remains well below the prediction of hyper-strong
  scattering, it varies by a factor of 2 to 4 on timescales of months
  at frequencies between 1.4 and 8 GHz (Spitler et al., in
  prep.). This suggests that local ``interstellar weather'' certainly
  plays a role and that nearby scattering screens also affect the
  observed emission, making the resulting ability to observe sources
  overall line-of-sight dependent, especially at lower
  frequencies. This is not unexpected given the properties of the
  turbulent interstellar medium in the Galactic Centre. Rather than
  dealing with a uniform single screen, it is likely that we see the
  effects of multiple finite screens. In this case, secondly, one
  expects a much shallower frequency dependence of the scattering time
  than the canonical $-4$ values (Cordes \& Lazio 2001).  This is
  indeed seen for high-DM pulsars (L\"ohmer et al.~2001, 2004), where
  the scattering index is typically around $\sim -3.5$ for large
  dispersion measures. Spitler et al.~(in prep.) find similar values
  for the magnetar. If this is indeed representative for a possible
  central pulsar or millisecond pulsar population, then the remaining
  scattering at 5, 14 or even 19 GHz would be underestimated in the
  analysis by Macquart et al.~(2010) or Dexter \& O'Leary (2014) by
  factors of 2.2, 3.7 or 4.3 respectively, when extrapolating from 1
  GHz. L\"ohmer et al.~(2001) measured even flatter frequency
  dependencies, which would make the discrepancy between real and
  estimated scattering times even larger. Unless more scatter
  broadening times in the Galactic centre are measured, this issue is
  difficult to settle. However, there is yet another, third effect
  that has been usually neglected in sensitivity calculations of
  pulsar surveys. As shown very recently by Lazarus et al.~(2015) for
  the P-ALFA survey, red noise present in pulsar search data due to
  radio interference (RFI), receiver gain fluctuations, and opacity
  variations of the atmosphere cause a significant decrease in
  sensitivity for pulsars with periods above 100 ms or so, when
  compared to the standard radiometer-based equation (see their
  Fig.~11). This would affect in particular a search for young
  pulsars, but also, of course, magnetars, which are nevertheless
  still easier to detect at high frequencies due to their much flatter
  flux density spectrum (Torne et al.~2015). This selection effect in
  particular would favour the detection of magnetars over that of
  normal, young pulsars and may explain in some respects the
  peculiarities of the current observational situation pointed out by
  Dexter \& O'Leary. The work by Lazarus et al.~demonstrates that the
  various selection effects are highly dependent on the individual
  surveys and that much more work is needed to understand the impact
  on the resulting search sensitivities.

  Finally, none of the previous high-frequencies surveys has, to our
  knowledge, applied a fully coherent acceleration search.  Such an
  acceleration search may be needed to account both for the movement
  of the pulsar around the central black hole, as well as for the
  presence of a binary companion. Indeed, due to the high stellar
  density, even exotic systems (e.g. MSP-stellar mass BH binary) may
  be expected (Faucher-Gigu{\`e}re \& Loeb 2011)\nocite{fl11}. An acceleration search is usually very
  computationally expensive, especially for long integration times as
  employed in the high frequency searches (e.g. by Macquart et
  al. 2010 or Eatough et al., in prep.), since the parameter range to
  be searched scales as $\propto T_{obs}^3$. The lack of such an
  acceleration search contributes especially as a selection effect to
  the present non-detection of fast-spinning pulsars.

  In order to model the selection effects (red noise, acceleration,
  scattering etc.) a more detailed study, taking also the orientation
  of the possible orbits and the change in acceleration into account,
  is needed. This is beyond the scope of this paper and will be
  presented elsewhere.  It is clear, however, that selection effects
  are not adequately modelled so far and that more work is required.}

We conclude that three scenarios are still possible: {\em (a)\/}~The
scattering seen for the Galactic Center magnetar is representative of
the inner parsec. In this case, the pulsar population may be dominated
by millisecond pulsars, for which this moderate scattering would still
have prevented their detection at previous search frequencies. Higher
frequency searches may therefore even allow the discovery and hence
the exploitation of millisecond pulsars orbiting \sgr\ (see also
Macquart \& Kanekar 2015)\nocite{mk15}.  {We note in passing
  that the discovery of a millisecond pulsar population may settle an
  ongoing debate about a possible excess of GeV gamma ray photons from
  the Galactic Centre. It is being discussed whether such an excess
  could arise from the presence of dark matter or a central population
  of unresolved young or millisecond pulsars (see e.g.~\cite{okkd15}
  and references therein). Any pulsar discovery in the Galactic Centre
  would make a dark matter discovery less likely.}  {\em (b)\/}~There
is a reduced number of pulsars in the Galactic Center region that is
consistent with selection effects. For example, the lack of dispersion
makes the discovery of unknown pulsars actually more difficult at high
frequencies as the signals are more difficult to be distinguished from
radio interference (see Eatough et al., in prep.), or {\em
  (c)\/}~PSR~J1745$-$2900 is indeed in front of a much more severe
scattering screen but the scattering properties for particular
line-of-sights are changing with time due to ``local weather''
effects, signs of which have been already detected (Spitler et al., in
prep.). In this case, search observations at even higher frequencies
are required and still promising.

Given that we cannot distinguish between these scenaria based on the
available data, high frequencies searches will continue. The use of
more sensitive instruments than available in the past, e.g., ALMA or
the Square Kilometre Array (SKA), may therefore lead to the discovery
of normal pulsars and even millisecond pulsars. In considering how
they can be used to measure the properties of \sgr, we will therefore
assume a variety of obtainable timing precisions. For details, we
refer the reader to Liu et al.\ (2012), who demonstrated possible
precision levels as a function of observing frequency for the SKA and
100-m class telescopes. {In their arguments, Liu et al.\ (2012)
  only considered normal pulsars and also assumed a hyper-strong
  scattering.  If the latter is not present as we have discussed
  above, millisecond pulsars may be detectable (although this may
  require proper acceleration searching).  Hence, for the discussion
  of the measurable effects, we will also allow for this possibility
  that a millisecond pulsars will be detected.

 There are a number of millisecond pulsars in globular clusters at
 distances that are signifantly larger than that of the Galactic
 centre. It is not uncommon to achieve a timing precision of about
 10\,$\mu$s for these distant sources. The exact precision depends
 foremostly on the strength of the pulsar signal and the sensitivity
 of the telescope, as well as the sharpness of some of the detetable
 profile features. If we need to go to high radio frequencies in order
 to beat interstellar scattering to see pulsars in the center of the
 Milky Way, the flux density decreases and timing precision decreases
 accordingly. This can be compensated by larger bandwidth or bigger
 telescopes. As shown in Eatough et al.~(2015), a timing precision of
 1\,$\mu$s should be routinely possible with the SKA, even at
 distances of the Galactic centre at higher frequencies. Such a
 precision is certainly more challenging with existing instruments.
 Overall, in order to cover all three plausible scenarios discussed
 above, we will assume, in the following, that a Galactic Center
 pulsar can be timed with a precision of 1, 10, and 100\,$\mu$s. As,
 in principle, only one pulsar is needed to extract the black-hole
 parameters, we consider this to be a useful range to demonstrate the
 effects that we can expect to measure.}


\subsection{Relativistic Orbital Effects}

\begin{figure}[t]
\begin{center}
  \includegraphics[height=7cm]{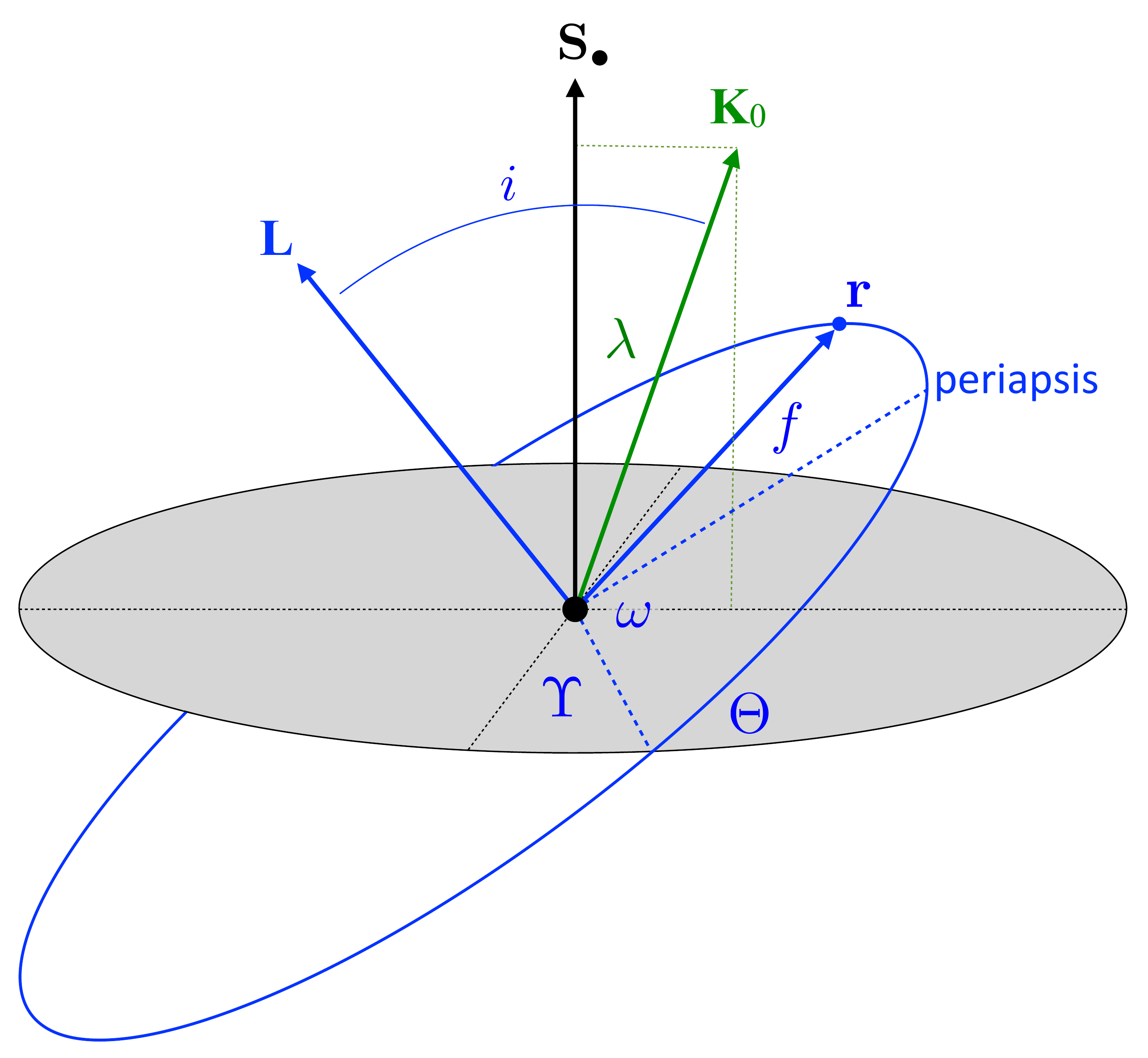}
\caption{Coordinate system and notation used in defining an orbit of a
  stellar-mass object around \sgr. The vector ${\bf S}_\bullet$
  denotes the spin of \sgr\ and ${\bf K}_0$ is line-of-sight unit
  vector pointing from the Earth to the black hole. The longitude of the
  periapsis of the orbit is $\omega$, the location of the ascending
  node is $\Upsilon$, and the inclination of the orbit with respect to
  the black-hole spin axis is $\Theta$. The angle $i$, between
    ${\bf K}_0$ and the orbital angular momentum ${\bf L}$, is the
    inclination of the orbit with respect to the observer.
\label{fig:psrbh}}
\end{center}
\end{figure}

In describing the orbit of a stellar-mass object around \sgr, we will
use the coordinate system and notation shown in
Figure~\ref{fig:psrbh}. In particular, we will denote by $m_*$ the
mass of the orbiting object and by $a$ and $e$ the semi-major axis
and eccentricity of its orbit. We will use the vector ${\bf S}_\bullet$ to
define the black-hole spin and the vector 
${\bf K}_0$ to denote the line-of-sight unit vector pointing from the Earth 
to the black hole. We will also denote the longitude of the periapsis of the
orbit {with respect to the equatorial plane of the black hole} by 
$\omega$, the location of the ascending node by $\Upsilon$,
and the inclination of the orbit with respect to the black-hole spin
axis by $\Theta$.

With these definitions, the Newtonian period of the orbit is
\begin{eqnarray}
P &=& 2\pi\left(\frac{a^3}{GM_\bullet}\right)^{1/2}
  \nonumber\\ 
  &=& 133.1\left(\frac{M_\bullet}{4\times 10^6\,M_\odot}\right)
  \left(\frac{ac^2}{GM_\bullet}\right)^{3/2}\mbox{s}\;.
\label{eq:period}
\end{eqnarray}

{Eccentric orbits of stars and pulsars precess on the orbital plane
(relativistic periapsis precession). The leading term comes from the
mass-monopole $M_\bullet$ and corresponds to the relativistic
precession of the Mercury orbit \citep{ein15}.
The advance of periapsis per orbit is $\Delta\omega = 2\pi k$, where 
\begin{equation}
  k = \frac{3}{1 - e^2} \, \left(\frac{2\pi GM_\bullet }{P c^3}\right)^{2/3}=
   \frac{3}{1 - e^2} \, \left(\frac{GM_\bullet}{c^2 a}\right)\;.
  \label{eq:k1pn}
\end{equation}
This corresponds to a characteristic timescale for this precession of
(see Merritt et al.\ 2010 for the definition, who denote this by $t_{\rm S}$)
\begin{eqnarray}
t_{\rm M}&\equiv&\frac{\pi P}{\Delta \omega}=\frac{P}{6}\,\frac{c^2
  a}{GM_\bullet}\left(1-e^2\right) \nonumber\\ &=&
22.18\left(\frac{M_\bullet}{4\times 10^6\,M_\odot}\right)
\left(\frac{ac^2}{GM_\bullet}\right)^{5/2}\left(1-e^2\right)\,\mbox{s}\;.
\label{eq:ts}
\end{eqnarray}
In this expression, we have neglected the small contributions of the
spin and of the quadrupole of the black hole.

Orbits with angular momenta that are not parallel to the spin ${\bf
  S}_\bullet$ of the black hole show a precession of the orbital
angular momentum around the ${\bf S}_\bullet$ direction due to frame
dragging (Lense-Thirring precession of the nodes).  The location of
the ascending node of the orbit, $\Upsilon$, advances per orbit by
\begin{equation}
  \Delta \Upsilon = \Omega_{\rm LT} P
\end{equation}
where
\begin{equation}
  \Omega_{\rm LT}\equiv \frac{8\pi^2}{(1-e^2)^{3/2}}
  \frac{GM_\bullet}{c^3 P^2}\chi
\end{equation}
is the Lense-Thirring frequency. 
The characteristic timescale for this process is (Merritt et al.\ 2010)
\begin{eqnarray}
t_{\rm J} &\equiv& \frac{\pi P}{\Delta \Upsilon}=\frac{P}{4\chi}
  \left[\frac{c^2a\left(1-e^2\right)}{G M_\bullet}\right]^{3/2}\nonumber\\
&=& 33.27\,\chi^{-1}\left(\frac{M_\bullet}{4\times 10^6\,M_\odot}\right)
\left(\frac{ac^2}{GM_\bullet}\right)^{3}\left(1-e^2\right)^{3/2}~\mbox{s}
\;.\nonumber\\
\label{eq:tj}
\end{eqnarray}

Finally, tilted orbits also precess because of the quadrupole moment
of the spacetime with a characteristic timescale (Merritt et
al.\ 2010)
\begin{eqnarray}
t_{\rm Q} &=& \frac{P}{3\vert q\vert}
  \left[\frac{c^2a\left(1-e^2\right)}{G M_\bullet}\right]^2\nonumber\\
&=& 44.35\vert q\vert^{-1}\left(\frac{M_\bullet}{4\times 10^6~M_\odot}\right)
\left(\frac{ac^2}{GM_\bullet}\right)^{7/2}\left(1-e^2\right)^{2}~\mbox{s}
\;.\nonumber\\
\label{eq:tq}
\end{eqnarray}
}
Figure~\ref{fig:timescales_stars} shows the characteristic timescales
of these relativistic orbital effects as a function of the semi-major
axes and orbital periods of the orbits. A number of additional
relativistic effects related to {time dilation and photon 
propagation (Shapiro delay)} can also be detected during timing
observations of pulsars. We will discuss these effects and their
dependence on the pulsar orbital parameters in \S\ref{sec:psr}.


\subsection{Optimal Orbital Parameters for Stars and Pulsars}

Performing tests of the no-hair theorem with orbits is hampered by a
number of astrophysical complexities caused by non-relativistic
phenomena that affect, in principle, the orbits. These included the
self-interaction between the stars in the stellar cluster (Merritt et
al.\ 2010; Sadeghian \& Will 2011), the hydrodynamic drag between the
stars and the accretion flow (Psaltis 2012), as well as stellar winds
and tidal deformations (Psaltis, Li, \& Loeb 2012). In order to
identify the orbital parameters of stars that are optimal for
performing the test of the no-hair theorem, we will first summarize
and combine the results of these studies.

\noindent {\em Interactions with Other Stars.---\/}Merritt et
al.\ (2010) and Sadeghian \& Will (2011) explored the decoherence of
the orbit of a star (or pulsar) due to Newtonian gravitational
interactions within the inner stellar cluster. They obtained an
approximate expression for the decoherence timescale given by
\begin{equation}
  t_{\rm N} = \frac{P}{q_*\sqrt{N(a)}} \;,
  \label{eq:tN}
\end{equation}
where $q_*\equiv m_*/M_\bullet$ is the average ratio of the mass of a
star (or compact object) in the cluster to that of \sgr, and $N(a)$ is
the number of stars inside the orbit of the object under
consideration.

Using equations~(\ref{eq:Na}), (\ref{eq:period}), and (\ref{eq:tN}), we
obtain for the decoherence timescale of orbits due to the self interaction
between objects in the stellar cluster
\begin{eqnarray}
t_{\rm N} &=& 2\pi \left(\frac{M_\bullet^2}{M_* m_*}\right)^{1/2}
  \left(\frac{a}{a_0}\right)^{-1/2}
  \left(\frac{a^3}{G M_\bullet}\right)^{1/2}\nonumber\\
&=& 12.6 \times 10^8
  \left(\frac{M_\bullet}{4.3\times 10^6 M_\odot}\right)^{3/2}
  \left(\frac{M_*}{10^6 M_\odot}\right)^{-1/2}\nonumber\\
&&\qquad
  \left(\frac{m_*}{M_\odot}\right)^{-1/2}
  \left(\frac{a_0}{1\,\mbox{pc}}\right)^{-1/2}
  \left(\frac{ac^2}{GM_\bullet}\right)\,\mbox{s}\;.
\end{eqnarray}

\begin{figure}[t]
\begin{center}
  \includegraphics[height=7cm,width=8cm]{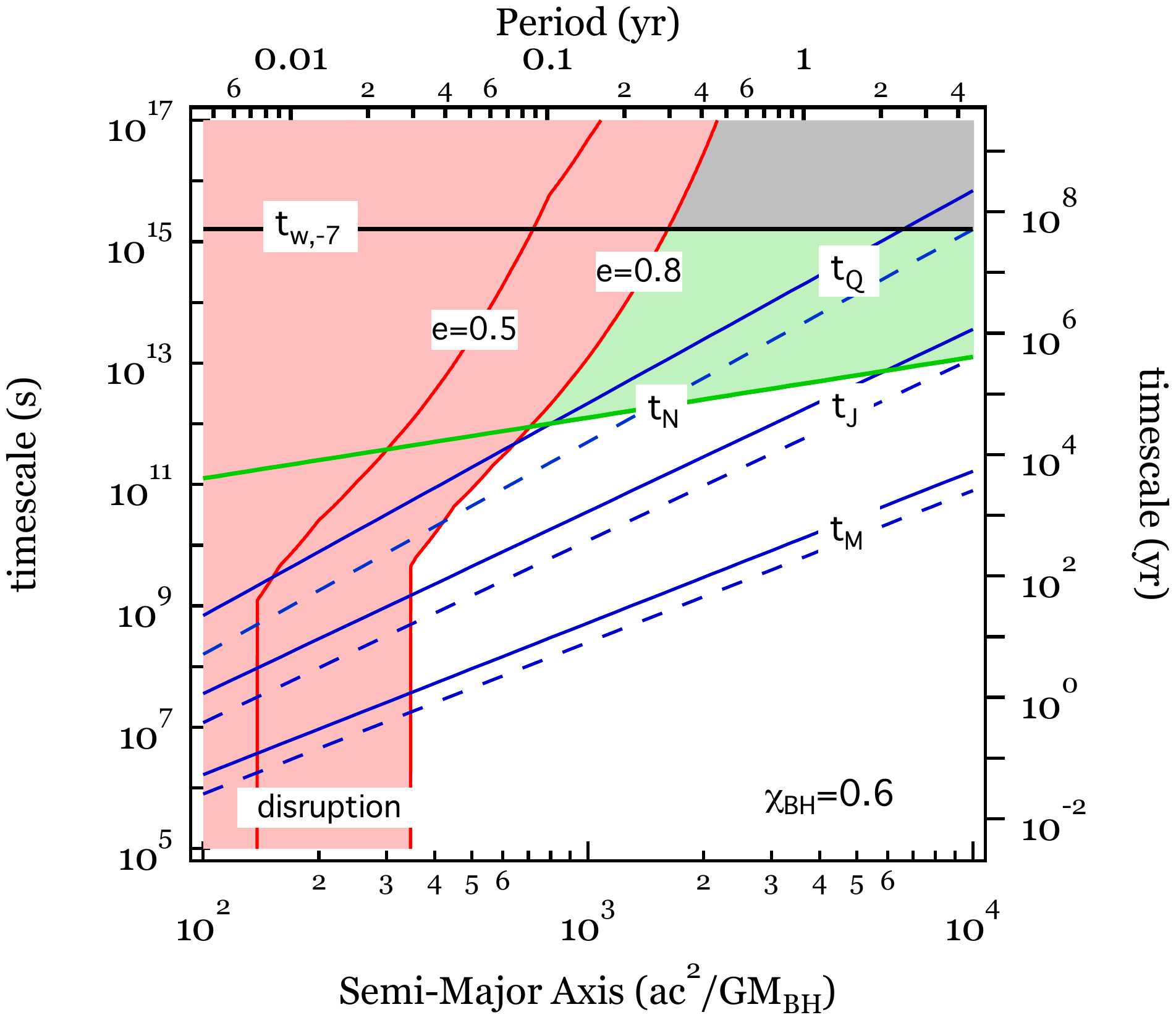}
\caption{Characteristic timescales for various relativistic and
  astrophysical effects that alter the orbits of stars around \sgr.
  The three blue lines correspond to the periapsis precession ($t_{\rm
   M}$), and orbital plane precession due to frame dragging ($t_{\rm
    J}$) and due to the quadrupole of the black hole ($t_{\rm Q}$),
  for orbits with eccentricities of $e=0.5$ (solid) and $e=0.8$
  (dashed), respectively; the spin of \sgr\ is set to $\chi=0.6$. The
  green line corresponds to the orbital decoherence timescale ($t_{\rm
    N}$) due to the interactions with other objects in the stellar
  cluster. The black curve ($t_{\rm w,-7}$) corresponds to the orbital
  evolution timescale due to the launching of a stellar wind. The red
  curves correspond to the orbital evolution due to the tidal
  dissipation of orbital angular momentum for two
  eccentricities. Stars in orbits with semi-major axes comparable to
  1000~gravitational radii are optimal targets for observing
  post-Schwarzschild relativistic effects.\label{fig:timescales_stars}
}
\end{center}
\end{figure}

Figure~\ref{fig:timescales_stars} compares the Newtonian decoherence
timescale with those of the three relativistic effects discussed in
\S2. For the parameters of \sgr\ and of the stellar cluster around it,
stars with orbital periods less than $\sim 1$\,yr are required in
order for the Newtonian interactions not to mask the orbital plane
precession due to frame dragging (see a more detailed discussion and
simulations in Merritt et al.\ 2010\nocite{2010PhRvD..81f2002M}).

\noindent {\em Hydrodynamic Interactions with the Accretion
  Flow.---\/} In Psaltis (2012), we investigated the changes in the
orbits of stars and pulsars caused by the hydrodynamic and
gravitational interactions between them and the accretion flow around
\sgr. For all cases of interest, we found that the hydrodynamic drag
is the dominant effect. However, as we will show below, even the
hydrodynamic drag is negligible for the orbital separations considered
here.

When a star of mass $m_*$ and radius $R_*$ plows through the accretion
flow of density $\rho$ with a relative velocity $u_{\rm rel}$, it
feels an effective acceleration equal to
\begin{equation}
a_{\rm d}=\frac{\pi R_*^2 \rho u_{\rm rel}^2}{m_*}\;.
\end{equation}
We can use this acceleration to define a characteristic timescale for
the change of the orbital parameters as
\begin{equation}
t_d\equiv \frac{u_{\rm rel}}{a_{\rm d}} = \frac{m_*}{\pi R_*^2 \rho u_{\rm rel}}\;.
\end{equation}
Setting the relative velocity equal to the orbital velocity of a circular
orbit, and the density of the accretion flow to
\begin{equation}
\rho=m_p n_e^0 \left(\frac{ac^2}{GM_\bullet}\right)^{-1.1}
\end{equation}
which has been inferred observationally (see discussion in Psaltis
2012), we obtain
\begin{eqnarray}
  t_{\rm d} &=& \frac{m_*}{\pi R_*^2 m_p n_0 c}
                \left(\frac{ac^2}{GM_\bullet}\right)^{1.6}\nonumber\\
            &=& 7.5\times 10^{15}
                \left(\frac{m_*}{10\,M_\odot}\right)
                \left(\frac{R_*}{10\,R_\odot}\right)^{-2}\nonumber\\
            &&\quad\left(\frac{n_e^0}{3.5\times 10^7\,\mbox{cm}^{-3}}\right)^{-1}
               \left(\frac{ac^2}{GM_\bullet}\right)^{1.6}\,\mbox{s}\;.
\end{eqnarray}
Here, $m_{\rm p}$ is the mass of the proton and we assumed for
simplicity that the orbit is circular. This timescale is significantly
larger than all other timescales shown in
Figure~\ref{fig:timescales_stars}.

\noindent {\em Stellar Winds.---\/}In Psaltis et al.\ (2012), we
explored the change in the orbital parameters of a stars due to the
launching of a wind that carries a fraction of the orbital energy and
angular momentum. The semi-major axis and the eccentricity of the
orbit change at a timescale comparable to $t_{\rm w} \equiv
m_*/\dot{M}_{\rm W}$, where $\dot{M}_{\rm w}$ is the rate of wind mass
loss, i.e.,
\begin{eqnarray}
t_{\rm w,-7} = 3.2\times 10^{15}
  \left(\frac{m_*}{10\;M_\odot}\right)
  \left(\frac{\vert\dot{M}_{\rm w}\vert}
             {10^{-7} M_\odot\,\mbox{yr}^{-1}}\right)^{-1}\,\mbox{s}\;,
\nonumber\\
\end{eqnarray}
where we have used the subscript ``$-7$'' to denote the exponent in the
wind mass loss rate. As shown in Figure~\ref{fig:timescales_stars},
the evolution of the stellar orbit due to the launching of a stellar
wind is always negligible compared to the effects of the Newtonian
interactions with the other stars in the cluster.

\noindent {\em Tidal Evolution.---\/}Finally, in Psaltis et
al.\ (2012), we also explored the evolution of a stellar orbit due to
the tidal dissipation of orbital energy during the periapsis passages.
Even though tidal dissipation does not cause a significant precession
in the orbit (see Sadeghian \& Will 2011), it leads to an evolution of
the semi-major axis that may be misinterpreted (due to the expected
low signal-to-noise ratio in the observations) as a change in the
projected orbital separation caused by orbital precession.

The characteristic timescale for orbital evolution due to tidal dissipation
is
\begin{eqnarray}
t_{\rm d}&\equiv& \frac{E}{\Delta E/\Delta t}\nonumber\\
&=&\frac{\pi R_{\rm *}}{c}
\left(\frac{G M_\bullet}{c^2 R_*}\right)^6
\left(\frac{m_*}{M_\bullet}\right)
\left(\frac{ac^2}{GM_\bullet}\right)^{13/2}(1-e)^6T_2^{-1}\nonumber\\
&=&0.98\times 10^{-4}\left(\frac{M_\bullet}{4.3\times 10^6~M_\odot}\right)^{5}
\left(\frac{R_*}{10~R_\odot}\right)^{-5}
\nonumber\\
&&\quad
\left(\frac{m_*}{10~M_\odot}\right)
\left(\frac{ac^2}{GM_\bullet}\right)^{13/2}\left(1-e\right)^{6}
T_2^{-1}(\eta)~\mbox{s}\;,
\end{eqnarray}
where the quantity $T_2(\eta)$ is defined and calculated in Psaltis et
al.\ (2012). Also, if the star at periastron reaches inside the tidal
radius
\begin{equation}
  R_{\rm t} = R_* \left(\frac{M_\bullet}{m_*}\right)^{1/3}\;, 
\end{equation}
it gets disrupted. Both these effects, for two different orbital
eccentricities, are shown in Figure~\ref{fig:timescales_stars}.

\noindent {\em Optimal parameters.---\/} Comparing the various
constraints shown in Figure~\ref{fig:timescales_stars} to the
characteristic timescales of the three relativistic effects allows us
to identify the optimal orbital parameters of stars and pulsars for
measuring the black-hole spin and for testing the no-hair theorem.

Using the orbital precession of stellar orbits to measure the spin of
\sgr\ simply requires sub-year orbital periods, for the effects of the
stellar perturbations to become negligible (as previously discussed in
Merritt et al.\ 2010). On the other hand, measuring the black-hole
quadrupole requires stars in much tighter orbits ($\lesssim 0.1$~yr),
for the stellar perturbations to be negligible, but with moderate
eccentricities ($e\lesssim 0.8$), for tidal effects to not interfere
with the measurements of the relativistic precessions (see also Will
2008).

For the case of pulsar timing, tidal effects do not alter the orbits
and, therefore, only stellar perturbations can limit our ability to
observe relativistic precessions. If we were to use pulsar timing to
measure the black-hole quadrupole by observing the pulsar orbital
plane precess, we would still be limited to using only rather tight
orbits ($\lesssim 0.1$\,yr). However, in defining the characteristic
timescale for quadrupole effects on the pulsar orbits
(eq.~[\ref{eq:tq}]), we have only considered the secular precession of
the orbit. The most promising way to extract the quadrupole moment
from timing observations is through the periodic effects in the
orbital motion of the pulsar caused by the quadrupolar structure of
the gravitational field of \sgr\ \citep{wk99,lwk+12}. This is not only
the case for the quadrupole but also for the relativistic precession
of the periapsis due to the mass monopole \citep{dd85} and the
precession of the orbit due to the frame dragging \citep{wex95}. (See
also the discussion in Ang{\'e}lil \& Saha 2014.) As argued by
\cite{lwk+12}, such unique periodic features in the timing of a pulsar
around \sgr\ provide a powerful handle to correct for external
perturbations. As we will demonstrate with mock data simulations in
\S\ref{sec:psr}, timing a pulsar only during a small number of
successive periapses passages is sufficient to measure both the spin
and the quadrupole moment of \sgr.


\section{Probing the spacetime of \sgr\ with Stars}
\label{sec:stars}



Astrometric and spectroscopic studies of stars in the near vicinity of
the black-hole horizon will allow detecting a number of relativistic
effects that depend on the spin and the quadrupole moment of the black
hole. Two of these effects will be the dominant ones (see discussion
in Angellil et al.\ 2010). The first is the precession of the
periapsis of an elliptical orbit on the orbital plane, which will lead
primarily to a very accurate measurement of the black-hole mass (see,
e.g., Weinberg et al.\ 2005). The second is the precession of
the orbital plane due to frame dragging, which depends on both the
spin and the quadrupole moment of the black hole. Measuring the latter
for two or more stars, will allow disentangling their effects on the
orbits and hence lead, in principle, to a test of the no-hair theorem
(Will 2008).

In the context of a stellar orbit around \sgr, we write the secular
rate of change of the location of its periapsis as (Merritt et al.\
2010)\footnote{Note, Merritt et al.\ (2010) use
$\varpi \equiv \omega + \Upsilon\cos\Theta$.}
\begin{equation}
  \dot{\omega} = \frac{\pi}{t_{\rm M}} -
                 \frac{3\pi}{t_{\rm J}} \cos\Theta -
                 \frac{\pi}{2t_{Q}}(1-5\cos^2\Theta)
  \label{eq:omdotSJQ}
\end{equation}
and the rate of precession of its orbital plane as
\begin{equation}
  \dot{\Upsilon} = \frac{\pi}{t_{\rm J}}-\frac{\pi}{t_{\rm Q}}\cos \Theta\;,
  \label{eq:UpsdotSJQ}
\end{equation}
where the various characteristic timescales were defined in \S2. A
change in either of these angles will correspond to an angular
displacement in the sky that will need to be measured astrometrically.
Because of the large lever arm of an eccentric orbit, the accuracy of
such a measurement will be determined by the ability of the
observations to detect changes in the position of the sky during
apoapsis passages. In other words, the two relevant quantities are the
total angular displacement of the apparent position of the apoapsis after $N$
orbits, i.e.,
\begin{eqnarray}
  \Delta\theta_{\rm apoapsis}
  &=&NP\dot{\omega} \frac{a(1+e)}{D}\cos i
\nonumber\\
&=&N\frac{1+e}{1-e^2}\left(\frac{GM_\bullet/c^2D}{5.12~\mu{\rm as}}\right)\cos i
\;  {\mu{\rm as}}
\bigg[ 96.8\nonumber\\
&&-192.8\chi\left(\frac{ac^2}{GM_\bullet}\right)^{-1/2}
  (1-e^2)^{-1/2}\cos \Theta\nonumber\\
  && -4.7\vert q\vert\left(\frac{ac^2}{GM_\bullet}\right)^{-1}
  (1-e^2)^{-1}(1-5\cos^2 \Theta)\bigg]\nonumber\\
\end{eqnarray}
and the total angular displacement in the apparent position of the line
of nodes, i.e.,
\begin{eqnarray}
  &&\Delta\theta_{\rm node}
    = NP\dot{\Upsilon} \frac{a(1+e)}{D}\cos i
    \nonumber\\
    &&= N\frac{1+e}{(1-e^2)^{3/2}}
        \left(\frac{ac^2}{GM_\bullet}\right)^{-1/2}
        \left(\frac{GM_\bullet/c^2D}{5.12\,\mu{\rm as}}\right)\cos i\; \mu{\rm as} 
   \nonumber\\
   &&\left[12.6\chi -9.4\vert q\vert\left(\frac{ac^2}{GM_\bullet}\right)^{-1/2}
     (1-e^2)^{-1/2}\cos\Theta\right]\;.
   \nonumber\\
\label{eq:deltathetanode}
\end{eqnarray}

For a star at an orbital separation of 1000\,$GM_\bullet/c^2$ with an
eccentricity of $e=0.8$ around a Kerr black hole that is maximally
spinning, tracing the orbit over 5~years ($N\simeq 38$) will lead to a
total angular displacement of the apparent position of the periastron
of $\Delta\theta_{\rm apoapsis}\simeq 8$~mas and of the apparent
position of the line of nodes of $\Delta\theta_{\rm node}\simeq
300~\mu$as. Observations with a future 30-m class telescope are
expected to have an accuracy of $\sim 500~\mu$as (Weinberg et
al.\ 2005), while simulations of tracing of stellar orbits with
GRAVITY suggest an accuracy of $\sim 10-200~\mu$as, depending on the
brightness of the star (Stone et al.\ 2012). Since our goal is to use
stellar observations to measure the precession of their orbital
planes, we will focus on future observations with GRAVITY and assume 
nominal uncertainties in the relative astrometric positions of
$\sigma_\theta=10~\mu$as and $\sigma_\theta=100~\mu$as. 

The uncertainties in the measurement of the black-hole properties will
be determined primarily by the ability of the experiment to measure
the advance of the apoapsis between orbits. If we approximate the
procedure as consisting of measuring the differential astrometric
location of the apoapsis once per orbital period with an uncertainty
$\sigma_\theta$, then the uncertainty in the inferred rate of
(apoapsis or orbital-plane) precession after $N$ measurements 
(i.e., orbits) that are equidistant in time will be (see Press et
al.\ 1992)
\begin{equation}
  \sigma_{\rm rate}^2 = \frac{\cal S}{{\cal S} {\cal S}_{tt}-{\cal S}_t^2}\;,
\end{equation}
where
\begin{eqnarray}
  {\cal S} &\equiv&\frac{N}{\sigma_\theta^2}\nonumber\\
  {\cal S}_t&\equiv&\sum_{i=1}^{N}\frac{(i-1)P}{\sigma_\theta^2}=
  \frac{N(N-1)}{2\sigma_\theta^2}P\nonumber\\
  {\cal S}_{tt}&\equiv&\sum_{i=1}^{N}\frac{(i-1)^2P^2}{\sigma_\theta^2}=
  \frac{N(2N^2-3N+1)}{6\sigma_\theta^2}P^2\;.\nonumber\\
\end{eqnarray}
For $N>>1$, the uncertainty in the total displacement after $N$ orbits
is
\begin{equation}
  \sigma_{\Delta\theta}=NP\sigma_{\rm rate}=\left(\frac{12}{N}\right)^{1/2}
  \sigma_\theta\;
  \label{eq:deltatheta}
\end{equation}
and the uncertainty in the inferred rate of precession is
\begin{equation}
  \sigma_{\rm rate}=\left(\frac{12}{N}\right)^{1/2}\frac{\sigma_\theta}{NP}\;.
\end{equation}

\begin{figure}[t]
\begin{center}
\includegraphics[height=7cm,width=8cm]{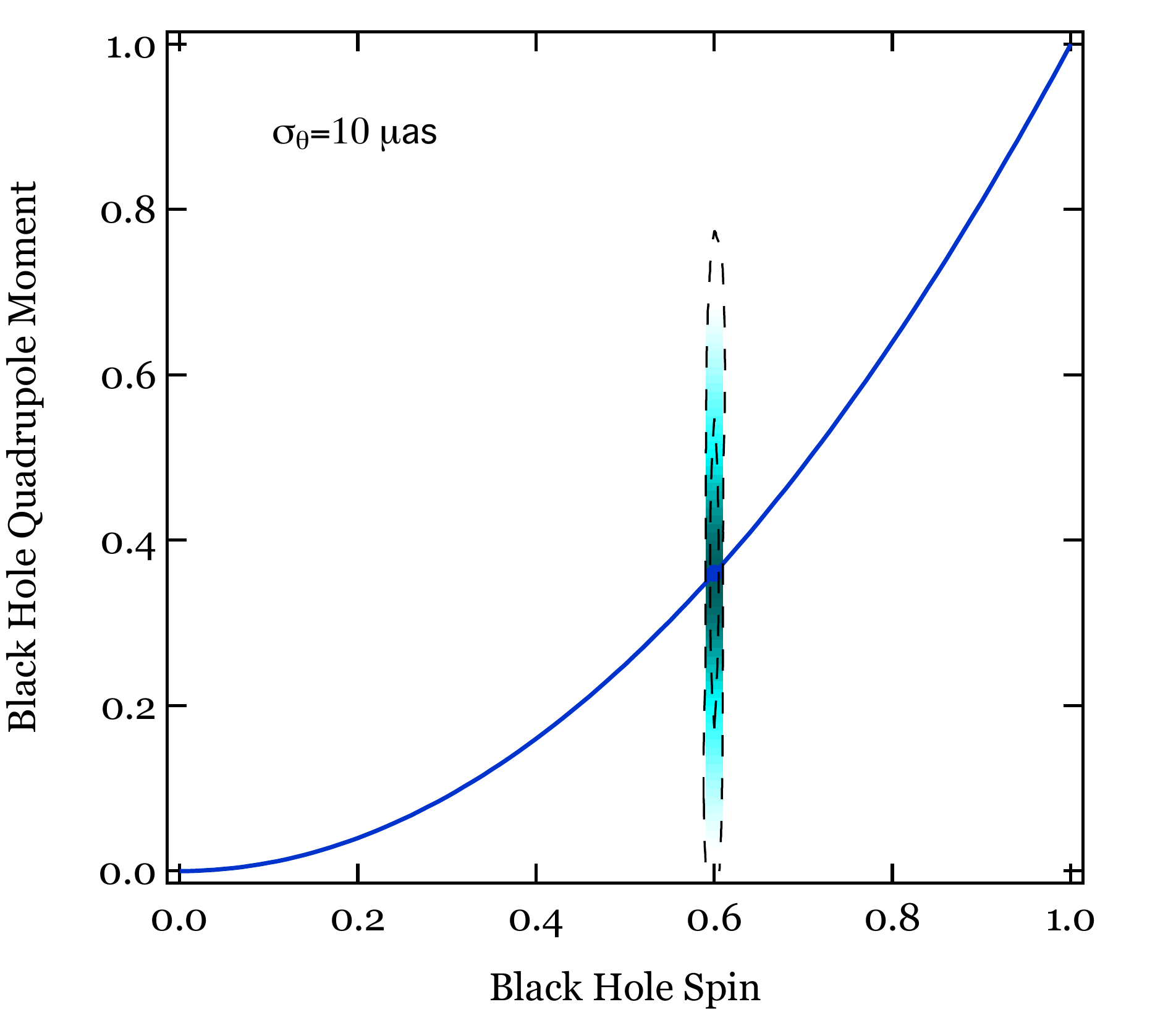}
\includegraphics[height=7cm,width=8cm]{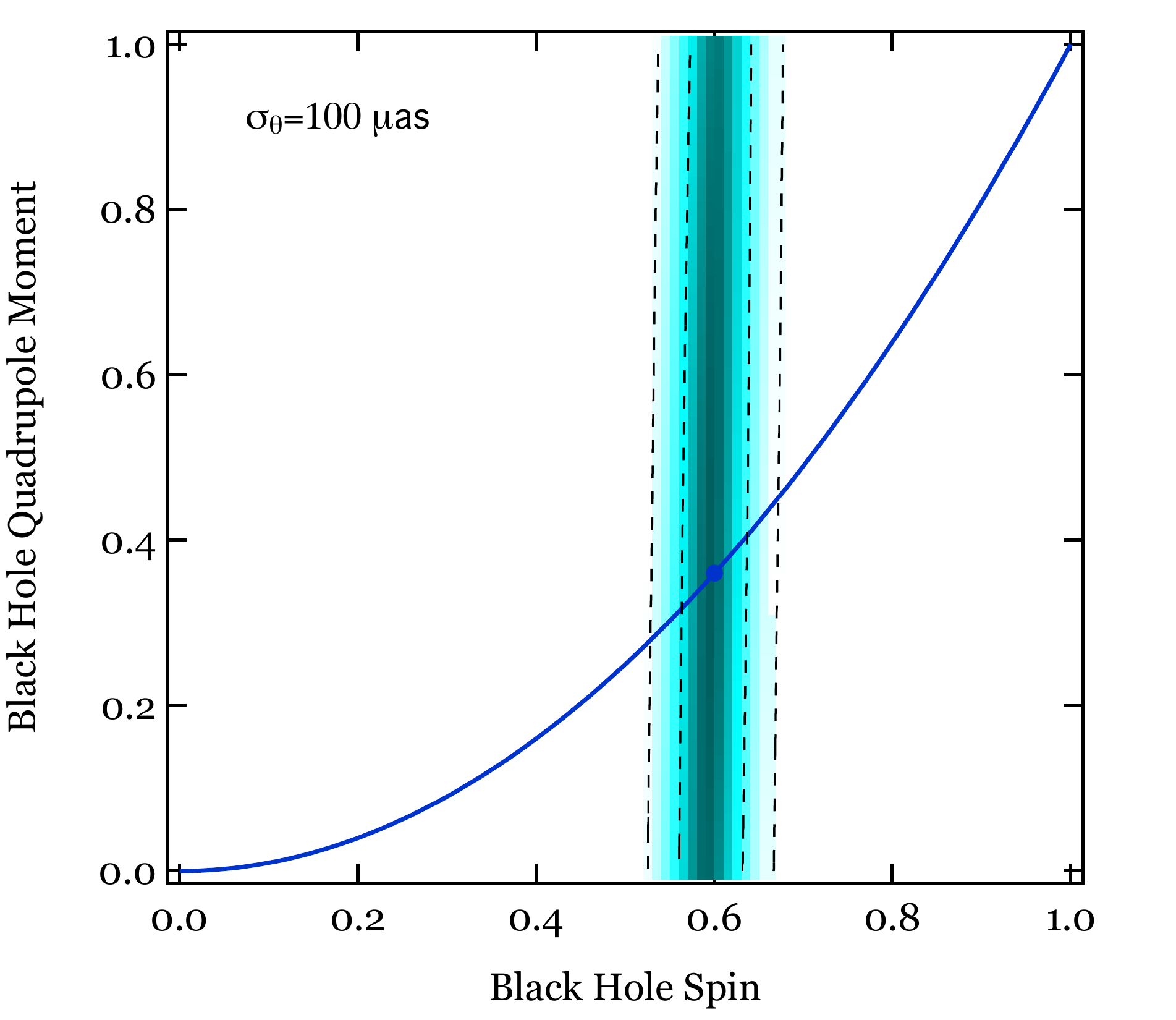}
\caption{The posterior likelihood of measuring the spin and quadrupole
  moment of \sgr\ by tracing the orbits of two stars with GRAVITY,
  assuming an astrometric precission of {\em (top)\/} $10~\mu$as and
  {\em (bottom)\/} $100~\mu$as. The dashed curves show the 68\% and
  95\% confidence limits, while the solid curve shows the expected
  relation between these two quantities in the Kerr metric.  The
  filled circle marks the assumed spin and quadrupole moment
  ($\chi=0.6$, $\vert q\vert=0.36$). The two stars are assumed to have
  orbital separations equal to 800 and 1000\,$GM_\bullet/c^2$ and
  eccentricities of 0.9 and 0.8, respectively. Even at these
  relatively small orbital separations, tracing the orbits of stars
  primarily measures the spin of the black hole, unless a very high
  level of astrometric precision is achieved. \label{fig:stars_nohair}
  }
\end{center}
\end{figure}

In order to illustrate the prospect of measuring the black-hole
properties using such measurements, we will assume that GRAVITY is
able to trace for $N=40$ orbits the trajectories of two stars with
orbital separations of 800 and $1000 GM/c^2$, with orbital
eccentricities of 0.9 and 0.8, respectively, and with the cosines of
all relevant orientations set to 0.5. We chose these orbits such that
the dynamical precession is faster than the other complicating
astrophysical effects we considered in \S2.4 (see Figure~2) and the
enclosed mass, angular momentum, and quadrupole moment of the stellar
cluster will not affect the measurements significantly.

Our goal is to estimate the posterior likelihood that a given
black-hole spin and quadrupole moment are consistent with the set of
measurements, i.e.,
\begin{equation}
  P(\chi,q|{\rm data})=P({\rm data}|\chi,q)P(\chi)P(q)\;.
\end{equation}
Here $P(\chi)$ and $P(q)$ are the priors on the black-hole spin and
quadrupole moment, which we take to be constant between zero and
unity. We also assume that the astrometric measurements for the two
stars are independent of each other (which will almost certaintly not
be true in reality), such that
\begin{equation}
  P({\rm data}|\chi,q) = \prod_{i=1}^2 P_{\rm apo}(i,{\rm data}|\chi,q)
  P_{\rm node}(i,{\rm data}|\chi,q)
\end{equation}
and $P_{\rm apo}(i,{\rm data}|\chi,q)$ and $P_{\rm node}(i,{\rm
data}|\chi,q)$ are the posterior likelihoods that a given black-hole
spin and quadrupole moment will generate the measurements for the
apoapsis and nodal precession of the $i$-th star, respectively. We
assume that the last two likelihoods for each star are Gaussian, with
centroids equal to the fiducial values that correspond to the orbits
of the stars around a Kerr black hole of spin $\chi=0.6$, and with
dispersions $\sigma_{\Delta\theta}$ given by
equation~(\ref{eq:deltatheta}) with $\sigma_\theta=10~\mu$as and
$\sigma_\theta=100~\mu$as.

Figure~\ref{fig:stars_nohair} shows the resulting posterior likelihood
over the black-hole mass and quadrupole moment, for the simulation
parrameters discussed above. As expected, even though using two stars
would allow us, in principle, to break the degeneracy between the spin
and the quadrupole moment of the black hole, in practice, GRAVITY
observations will be able to predominantly measure the spin of the
black hole. This, of course, can be performed even with following
the orbit of a single star. In that case, using
equation~(\ref{eq:deltathetanode}), we can estimate the accuracy to
which GRAVITY observations will lead to a measurement of the
black-hole spin as
\begin{eqnarray}
  \sigma_\chi &\sim&  0.064\left(\frac{\sigma_\theta}{100\,\mu{\rm as}}\right)
  \left(\frac{N}{40}\right)^{-3/2}
  \left(\frac{ac^2/GM_\bullet}{1000}\right)^{1/2} \nonumber\\ &&
  \left(\frac{GM_\bullet/c^2D}{5.1\,\mu{as}}\right)^{-1}
  \left[\frac{(1-e)(1-e^2)^{1/2}}{0.12}\right]
  \left(\frac{\cos i}{0.5}\right)^{-1}.\nonumber\\&&
\end{eqnarray}

In this last expression, we have neglected the correlated
uncertainties between the measurement of the spin of the black hole
and of the orientation of the orbit with respect to the
spin. Nevertheless, our estimates here show the ability of astrometric
tracing of stellar orbits with GRAVITY to determine the spin of \sgr\
and to provide an independent probe that will allow us to control and
quantify possible systematic effects in the measurement.


\section{Probing the spacetime of \sgr\ with a pulsar}
\label{sec:psr}



As demonstrated by \cite{wk99} and \cite{lwk+12}, a single pulsar
orbiting \sgr\ at similar distances as discussed for stellar orbits
earlier will allow us to extract the relevant black-hole parameters
with high precision, even if only a moderate timing precision {can be
achieved}. In order to gauge the feasibility of such an experiment, we
are, of course, limited by two major uncertainties. One is the
existence of detectable and timeable pulsars in appropriate distances
to the central black hole, and the other is the impact of potential
perturbations to the pulsar orbits due to external effects.

We addressed the first source of uncertainty in \S2.3.  In order to
address the second, we expand here on the earlier work by \cite{wk99}
and \cite{lwk+12}, who presented the fundamental recipe of this
experiment. The work presented here goes further, providing a major
step towards the development of a timing formula that can be used to
exploit the pulsars once they are discovered. In particular, it makes
use of a timing model that consistently includes the periodic spin
contributions derived in \cite{wex95}, where the orbital motion in the
reference frame of the black hole (Fig.~\ref{fig:psrbh}) is described
by the following quasi-Keplerian parametrization
\begin{eqnarray}
  n(t - t_0) &=& u - e_t \sin u \;, \label{eq:Kepler}\\
  f          &=& 2\arctan\left[
                 \left(\frac{1 + e_\varphi}{1 - e_\varphi}\right)^{1/2}
                 \tan\frac{u}{2} \right] \;, \\
  r          &=& a(1 - e_r \cos u) \;, \\
  \varphi    &=& \omega_0  + (1 + k) f \;, \\
  \Upsilon   &=& \Upsilon_0 + w (f + e \sin f) \;. \label{eq:Ups}
\end{eqnarray}
The orbital frequency $n$ is related to the orbital period $P$ via $n
= 2\pi/P$. The angle $\omega_0$ gives the location of the periapsis at
$t = t_0$. The three eccentricities $e_t$, $e_\varphi$ and $e_r$ are
different from each other only at the first post-Newtonian (pN) level,
and the quantities $k$ and $w$ are of 1pN and 1.5pN order,
respectively.

A comment on the practical use of above equations: The parametrization of 
the orbital motion in
equations~(\ref{eq:Kepler})--(\ref{eq:Ups}) represents a simple
extension of the elegant quasi-Keplerian solution of the 1pN two-body
problem, found by \cite{dd85}. The latter is the basis of the DD
timing model \citep{dd86}, which is implemented in TEMPO, the standard
software for pulsar timing\footnote{There are two TEMPO versions in
use, TEMPO (http://tempo.sourceforge.net/) and
TEMPO2 \citep{hem06}. For our simulations we used a modified version
of TEMPO.}. For this reason, we could easily extend TEMPO to include
spin-orbit and, as we discuss later, quadrupole effects. This
modified TEMPO version forms the basis of our mock data
simulations. 

The location of the periapsis at a time $t$ is given by
\begin{equation}
  \omega = \omega_0 + k f \;.
\end{equation}
Consequently the advance of periapsis is linear in the true anomaly,
and, therefore, non-linear in time. For highly eccentric orbits, the
advance of the periapsis is clearly faster when the pulsar is near the
central black hole. The orbital averaged precession rate is given by
\begin{equation}
  \langle\dot{\omega}\rangle = k n \;.
\end{equation}
As evident from equation~(\ref{eq:Ups}), a similar behavior comes with
the periodic spin contributions, where the orbital averaged precession
of the nodes (Lense-Thirring precession) is given by
\begin{equation}
  \langle\dot{\Upsilon}\rangle = w n \;.
\end{equation}
Both, $k$ and $w$ are free parameters of the timing model. For a given
theory of gravity, they depend on the Keplerian parameters of the
pulsar orbit and on the mass and the spin of the central black hole.

At this point it is important to note that $\omega$ is not the
longitude of periapsis that enters the pulsar-timing model directly. The timing
model makes use of the longitude of periapsis with respect to the plane
of the sky, and we denote it by $\tilde\omega$. Its relation to 
$\omega$ depends on the orientation of the black-hole spin with respect
to the observer and can be found in \cite{wk99}.

In the following, we first discuss the various relativistic effects
that allow us to use pulsar timing in order to measure the mass, the
spin, and the quadrupole moment of the central black hole. Then, using
the new timing model, we show in mock data simulations that it is
sufficient to time the pulsar when it moves near \sgr\, in order to
determine the mass, spin, and quadrupole moment of the black hole. As
a consequence, the no-hair theorem test with a pulsar turns out to be
fairly robust against external perturbations. Furthermore, we
investigate the possibility to fully determine the spatial orientation
of the \sgr\ spin and give an estimate for a distance measurement
from timing.


\subsection{Mass Determination}
\label{subsec:psr.mass}

It is well known that the measurement of post-Keplerian (PK)
parameters in binary pulsars can provide highly accurate measurements
of the masses of the system \citep{lk04}. The same can be expected for
a pulsar in orbit around \sgr, where the situation is insofar
different as the pulsar is like a test particle, whose mass is
negligible in comparison to the companion's mass, i.e., the 4.3
million solar masses of \sgr. In such a situation the measurement of a
single PK parameter allows the determination of the mass of \sgr, once
a theory of gravity is assumed. The measurement of a second PK
parameter already allows for a consistency check, since the inferred
mass should agree with the one from the first PK parameter \citep{lwk+12}.

In the following, we quickly summarize the most important
relativistic effects and their leading order expression within GR (see
Lorimer \& Kramer 2004\nocite{lk04}, and references therein for
details).
\begin{itemize}
\item {The advance of periapsis per orbit is $\Delta\omega = 2\pi k$, where 
$k$ was defined in equation~(\ref{eq:k1pn}). For a fast spinning black
hole, $k$ can have a significant contribution from frame-dragging
effects, as we will discuss in more detail below.}
\item The time dilation (Einstein delay) has an amplitude of
\begin{equation}
  \gamma_{\rm E} = 2\frac{e}{n}\left(\frac{GM_\bullet}{c^3}n\right)^{2/3}\;. 
  \label{eq:gam1pn}
\end{equation}
\item The signal propagation delay (Shapiro delay), which is proportional to the black hole mass, reads as a function of the true anomaly $f$ as
\begin{equation}
  \Delta_{\rm S} = \frac{2GM_\bullet}{c^3} \, \ln\left[\frac{1 +
    e \cos f}{1 - \sin i\sin(\tilde\omega + f)}\right] \;.
\end{equation}
\end{itemize}
A measurement of the Shapiro delay, simultaneously gives $M_\bullet$
and $\sin i$ (see Figure~\ref{fig:psrbh} for the definition of the
inclination angle $i$). The latter is connected to $M_\bullet$ via the
so-called mass function
\begin{equation}
  \sin i = n x \left(\frac{GM_\bullet}{c^3}n\right)^{-1/3} \;. 
\end{equation}
The Keplerian parameter $x \equiv a \sin i/c$ is the projected 
semi-major axis of the
pulsar orbit and can be measured with high precision from pulsar
timing. Consequently, there are two ways to extract the mass of \sgr\
from a measurement of the Shapiro delay.

Based on a consistent covariance analysis using mock data
simulations, \cite{lwk+12} demonstrated that a pulsar in a close orbit
($P_b \sim 1$~yr) around \sgr\ allows for a very precise determination
of its mass, $M_\bullet$. Even a moderate timing precision can lead to
a $\sim10^{-5}$ precision for $M_\bullet$, since there are various
relativistic effects that can be utilized for mass determination. The
simulations in \cite{lwk+12}, however, were based on the assumption
that the pulsar can be timed continuously over several years, covering
at least a few full orbits.  Later in this section, we relax this
assumption and allow for a situation where, due to external
perturbation, only the timing data near the black hole can be used in
a phase-connected solution. As it turns out, in particular the
Einstein and the Shapiro delay provide a robust determination of
$M_\bullet$ that is only weakly affected by external perturbations
or uncertainties in our knowledge of the spin of \sgr.


\subsection{Frame Dragging and Spin Measurement}

{The dragging of inertial frames by the rotation of the black hole
affects the precession of the periapsis of the orbit. Indeed, beyond
leading order, equation~(\ref{eq:k1pn}) would include a Lense-Thirring
contribution $k_{\rm LT}$, where
\begin{equation}
  k_{\rm LT} = -3 \frac{\Omega_{\rm LT}}{n} \cos\Theta
\end{equation}
(see also equation~[\ref{eq:omdotSJQ}]). An independent measurement of
the mass $M_\bullet$, for instance through the Shapiro delay, could
then be used to compute $\chi\cos\Theta$, within GR}, where
$\chi\cos\Theta \le 1$ is required by the cosmic censorship
conjecture.

The most prominent effect of frame dragging in the pulsar motion is
the Lense-Thirring precession of the orbital plane. In GR, the nodes
of the orbit precess at an averaged rate of
\begin{equation}
  \langle\dot\Upsilon\rangle = w n = \Omega_{\rm LT} \;,
\end{equation}
(see also equation~[\ref{eq:UpsdotSJQ}]). Although $w$ is a small
quantity ($\sim 10^{-4}\,\chi$ for a 0.1\,yr orbit), given the large
size of a pulsar orbit around \sgr, it has a tremendous impact on the
timing residuals. In fact, for orbits $\lesssim 1$~yr, it leads to a
large change in the projected semi-major axis, $x$, giving rise to a
significant time evolution of $x$, that can be measured as a first
derivative $\dot x$ and even a second derivative $\ddot x$. Furthermore, 
it also leads to an observable second derivative in the advance of the 
periapsis, $\ddot{\tilde\omega}$ (see Liu et al.\ 2012\nocite{lwk+12} 
for details).

\begin{figure}[t]
\begin{center}
\includegraphics[height=6cm]{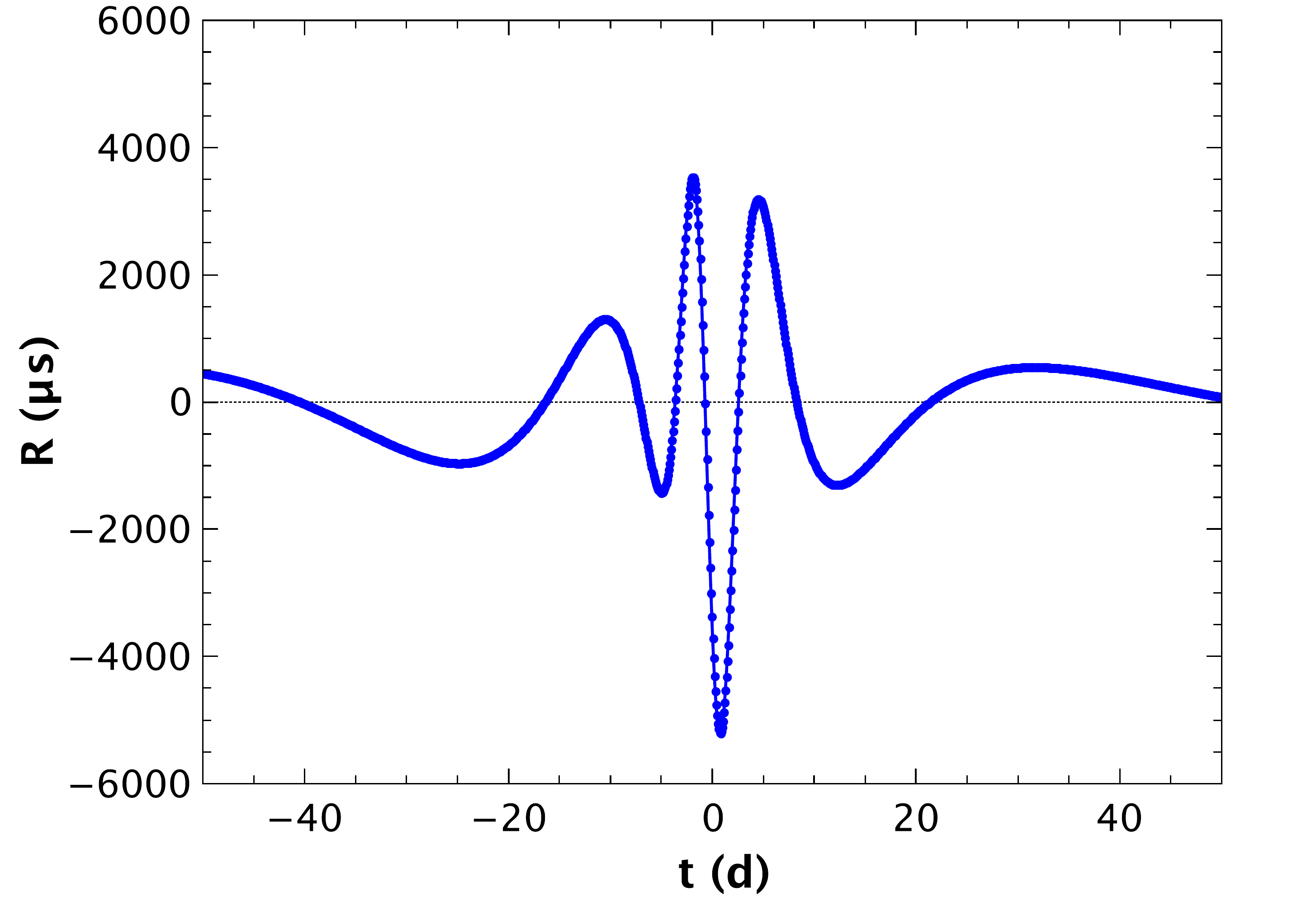}
\caption{Timing residuals near periapsis passage ($t=0$) for a pulsar in a
highly eccentric ($e=0.9$) 3\,yr orbit around \sgr, when frame
dragging effects have not been taken into account when fiting the
pulsar orbit. We assume a Kerr spacetime with spin parameter $\chi =
0.6$. Concerning the orientation of the black hole, we used $\Theta =
60^\circ$, $\Upsilon_0 = \omega_0 = 45^\circ$, and $\lambda =
55^\circ$. The last value is motivated by \cite{pnf+15}. The dense
timing campaign covers only one year around periapsis. Still, even
after fitting the full DD model and allowing for a secular precession
of the orbital plane, the frame-dragging (spin-orbit) contribution
gives rise to a strong characteristic feature in the timing residuals.
\label{fig:res_so_3yr_09}}
\end{center}
\end{figure}

As can be seen from equation~(\ref{eq:Ups}), the location of the
ascending node, $\Upsilon$, advances non-linearly in $f$ and $t$,
which we will exploit for the first time in this paper. Instead of
using only the secular contributions $\dot x$, $\ddot x$, and
$\ddot{\tilde\omega}$ to model the changes in the orbit due to 
Lense-Thirring precession, we implement the full model
equations~(\ref{eq:Kepler})--(\ref{eq:Ups}), therefore also accounting
for the periodic contributions. The latter is of particular
importance, if the pulsar is in a highly eccentric orbit. Morover,
this will turn out to be extremely valuable in the presence of
external perturbations. In fact, \cite{lwk+12} have already argued
that these distinctive near-periapsis contributions can be used to
differentiate between frame dragging by the black hole and external
contributions, which are more likely to affect the pulsar's motion
near the apoapsis (see also Ang\'elil \& Saha
2014\nocite{as14}). Figure~\ref{fig:res_so_3yr_09} illustrates such
Lense-Thirring contributions to the pulsar timing residuals for a
highly-eccentric ($e = 0.9$), wide ($P_b = 3$\,yr) pulsar orbit. From
Figure~\ref{fig:res_so_3yr_09} it is already clear that, from a single
periapsis passage that is covered by a dense observing campaign, we
can already infer relevant constraints on the \sgr\ spin. We will
present more detailed conclusions in the subsection on mock data
simulations below.

\subsubsection{Mock data simulations}

{While from the above discussion it is already obvious that the 
Lense-Thirring drag of the rotating black hole leads to a characteristic 
signal in the timing data, which ultimately allows the determination of 
the spin, we still need to address the question of spin measurement in a 
more quantitative way. For this reason, we have conducted extensive mock 
data simulations, based on the potential timing capabilities discussed in 
Section~\ref{sec:GCpsrs}.} The simulated
TOAs were fitted with a timing model based on the equations
of motion (\ref{eq:Kepler})--(\ref{eq:Ups}), that also include the
relativistic effects discussed in Section~\ref{subsec:psr.mass}. By this our simulations are based on a timing model
that accounts for all the relevant effects to leading order. This
model has been implemented in a timing software package, which is
based on TEMPO and has been optimized for the timing analysis of
Galactic Center pulsars. It allows for a fully phase-connected timing
solution, providing a consistent parameter estimation. This is similar to 
the analysis presented in \cite{lwk+12};
however, in addition, it properly accounts for the prominent
near-periapsis features in the residuals, caused by the Lense-Thirring
effect. As discussed above, this is of particular importance, if there
are external perturbations to the pulsar orbit, moreover it also helps in
determining the black hole spin on a shorter observing time-span,
during which the second derivatives $\ddot x$ and $\ddot{\tilde\omega}$ 
are not
well measured. The latter is important in case the pulsar is only
visible for a limited period of time, which will be the case for a
pulsar at the Galactic Center, as discussed in \S~\ref{sec:GCpsrs}.

\begin{figure}[t]
\begin{center}
\includegraphics[height=6cm]{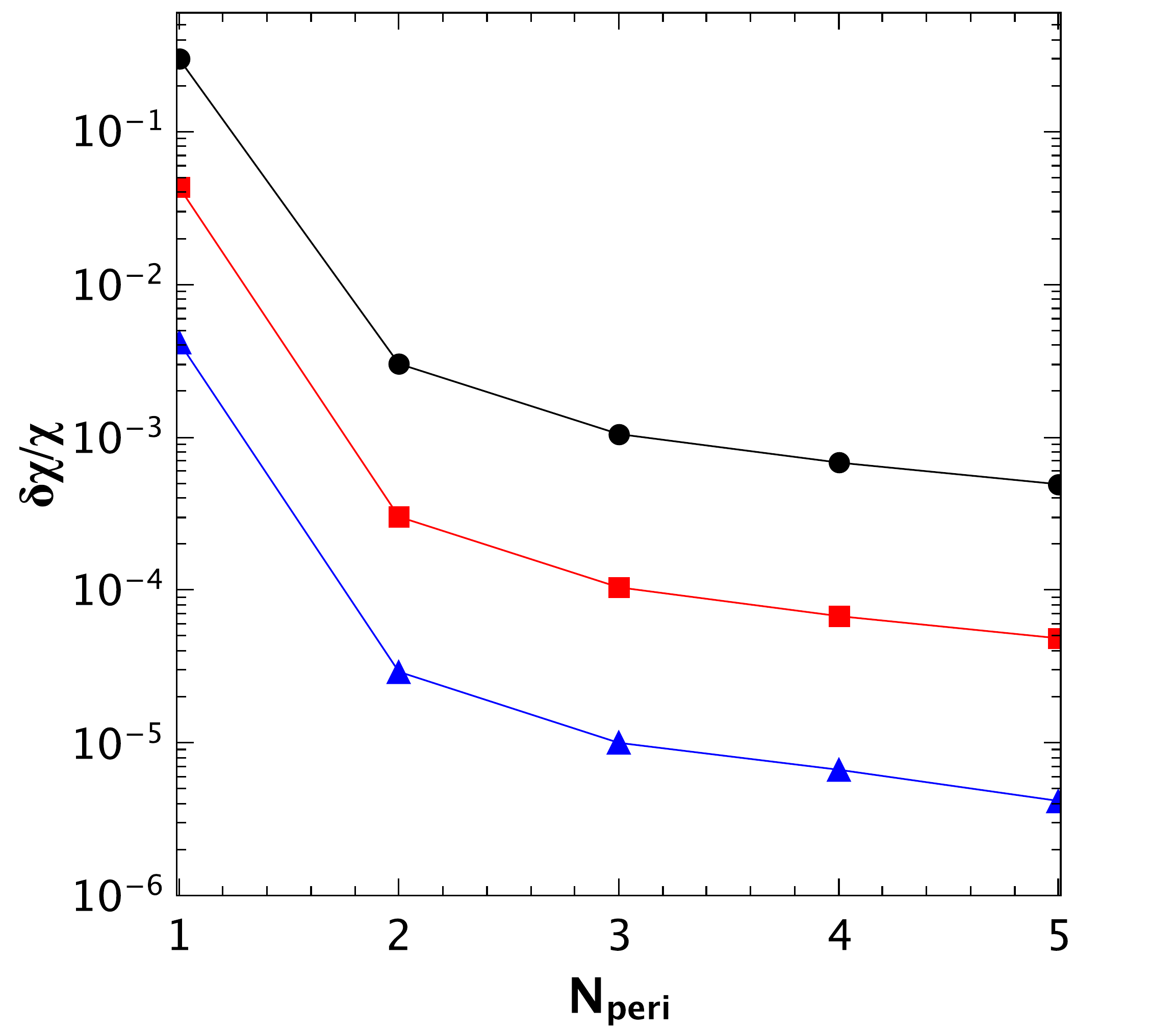}
\caption{Fractional measurement precision (2-$\sigma$) for the spin parameter
$\chi$ as a function of periapsis passages, based on a dense timing
campaign. We have use the following values for the various parameters:
$P_b = 0.5$\,yr, $e = 0.8$, $\chi = 0.6$, $\sigma_{\rm TOA} = 1\,\mu$s
(blue), 10\,$\mu$s (red), 100\,$\mu$s (black). The orientation of the
spin is taken as in Figure~\ref{fig:res_so_3yr_09}. We assumed a daily
timing campaign with three TOAs per session.}
\label{fig:dchi}
\end{center}
\end{figure}

Figure~\ref{fig:dchi} presents the measurement precision of the
dimensionless spin parameter $\chi$ as a function of the observed
number of periapsis passages for a pulsar in a 0.5\,yr eccentric ($e =
0.8$) orbit. We have simulated a dense observing campaign where one
obtains three precision TOAs per day. Simulations have been conducted
for three different TOA uncertainties: 1, 10, and 100\,$\mu$s,
corresponding to our discussion in Section~\ref{sec:GCpsrs}. As is
evident from Figure~\ref{fig:dchi}, for all these TOA uncertainties
we should be able to measure the spin of \sgr\ with very high
precision, even if the observing time-span covers only a few periapsis
passages. We have to keep in mind though, that the secular precession
of the orbit, which is used in Figure~\ref{fig:dchi} to determine the
spin, also has a contribution from the quadrupole moment of the black
hole. For a Kerr black hole and the orbits considered here, this
contribution is considerably smaller than the spin contribution
(cf.~Figure~\ref{fig:timescales_stars}). Nevertheless, for a $10^{-3}$
precision (or better) in the spin measurement within a no-hair-theorem
test, we need to account for the quadrupole contribution without a
priori assuming a fixed relation between spin and quadrupole
moment.

As already argued in \cite{wk99} and demonstrated in detail
in \cite{lwk+12}, we have a good handle on the quadrupole from the
characteristic periodic features in the timing residuals. In a
consistent analysis we can, therefore, measure the quadrupole moment
and simultaneously account for its contribution to the spin
determination based on the secular precession of the pulsar
orbit. Moreover, as we will discuss below in the subsection on the
quadrupole measurement, the periodic residuals caused by the
quadrupole moment have a signature that is quite different from the
signature of the periodic spin contributions, which further helps to
separate spin and quadrupole effects in the orbital motion.

\begin{figure}[t]
\begin{center}
\includegraphics[height=6cm]{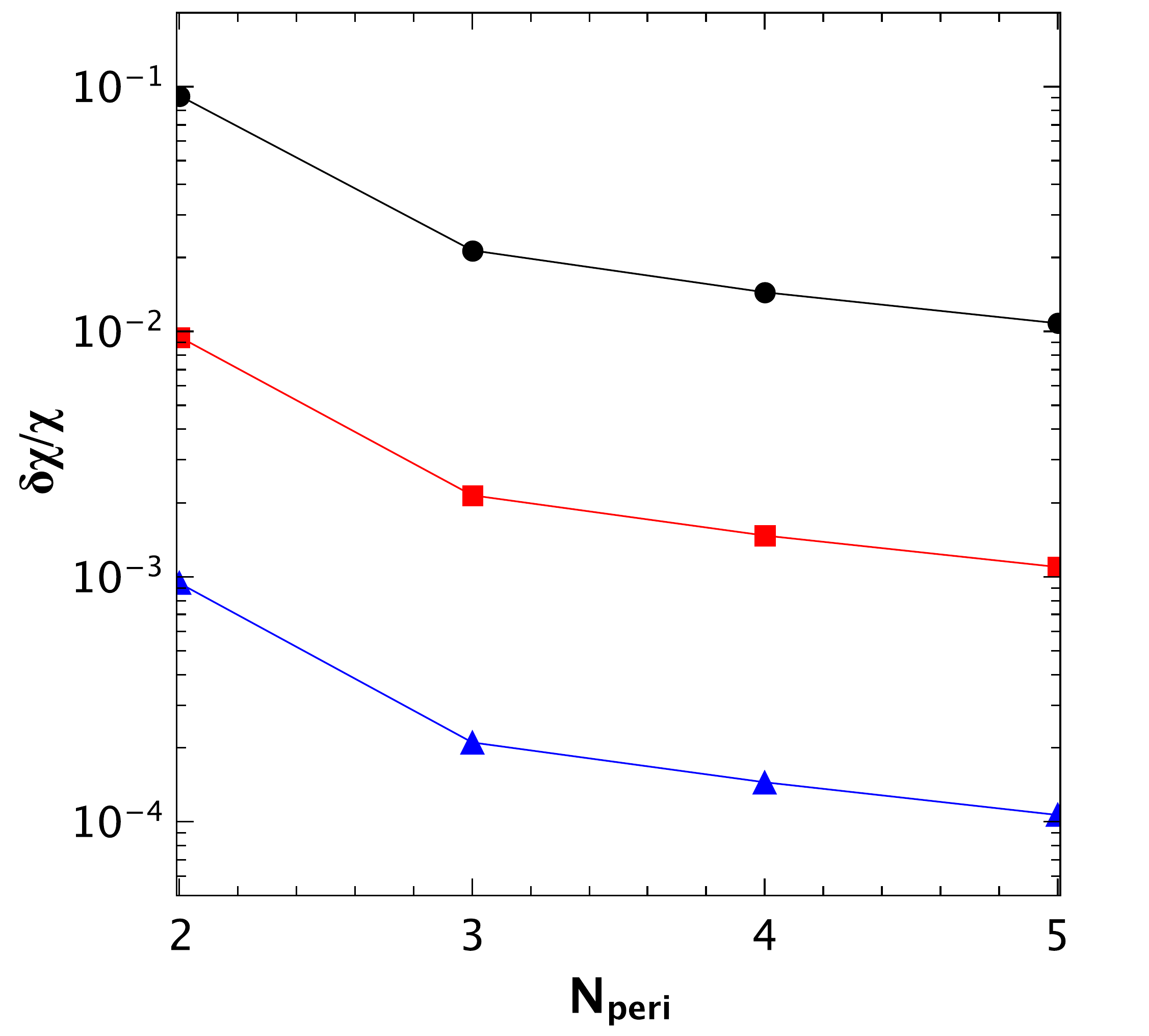}
\caption{Fractional measurement precision (2-$\sigma$) for the spin parameter
$\chi$ as a function of periapsis passages for a pulsar orbit that
suffers external perturbations and for which only TOAs near periapsis
can be used for parameter fitting (details in the text). The various
parameters are similar to those in Figure~\ref{fig:dchi}.
\label{fig:dchi_p}}
\end{center}
\end{figure}

In an ideal situation, the pulsar's motion around \sgr\ will only be
affected by the gravitational field of the black hole, as we have
assumed in the previous simulations. However, if the orbital motion of
the pulsar is exposed to external perturbations, for instance by a
nearby mass distribution due to stars or dark matter, then the orbit
might show an additional precession, which a priori cannot be 
quantified (see Merritt et al.~2010 and discussion in 
\S~\ref{sec:gen}). This is expected to be of
particular importance around apoapsis, where the pulsar spends most of
its time, and where the gravitational effects from the black hole are
weaker. In such a case, all the information on the black hole spin has
to come from the expectedly dominating spin effects near the
periapsis. We have modeled such a situation in our simulations by
taking only TOAs during the periapsis passages (time interval of only
$\pm 0.05\,P$ around periapsis), The estimated measurement precision
for the spin of \sgr\ is plotted in Figure~\ref{fig:dchi_p}. While the
measurement precision is weaker than in Figure~\ref{fig:dchi}, it is
obvious that it will still be possible to measure the spin of \sgr\
with high precision just based on the characteristic Lense-Thirring
signal near periapsis (cf.~Figure~\ref{fig:res_so_3yr_09}).

Finally, if we are only able to observe a single periapsis passage, of
a wide but highly eccentric orbit, like the one in
Figure~\ref{fig:res_so_3yr_09}, a complete spin measurement might be
out of range, due to the strong correlations with other timing
parameters. Nevertheless, as our simulations show, we should still be
able to get precise constraints on different spin-projections, like
$\chi\cos\Theta$ and $\chi\sin\lambda$, similar to the situation
in \cite{lwk+12}, if for instance none of the higher derivatives in
$\tilde\omega$ and $x$ can be measured.


\subsubsection{Determining the spatial orientation of the \sgr\ spin}

\begin{figure}[t]
\begin{center}
\includegraphics[height=6cm]{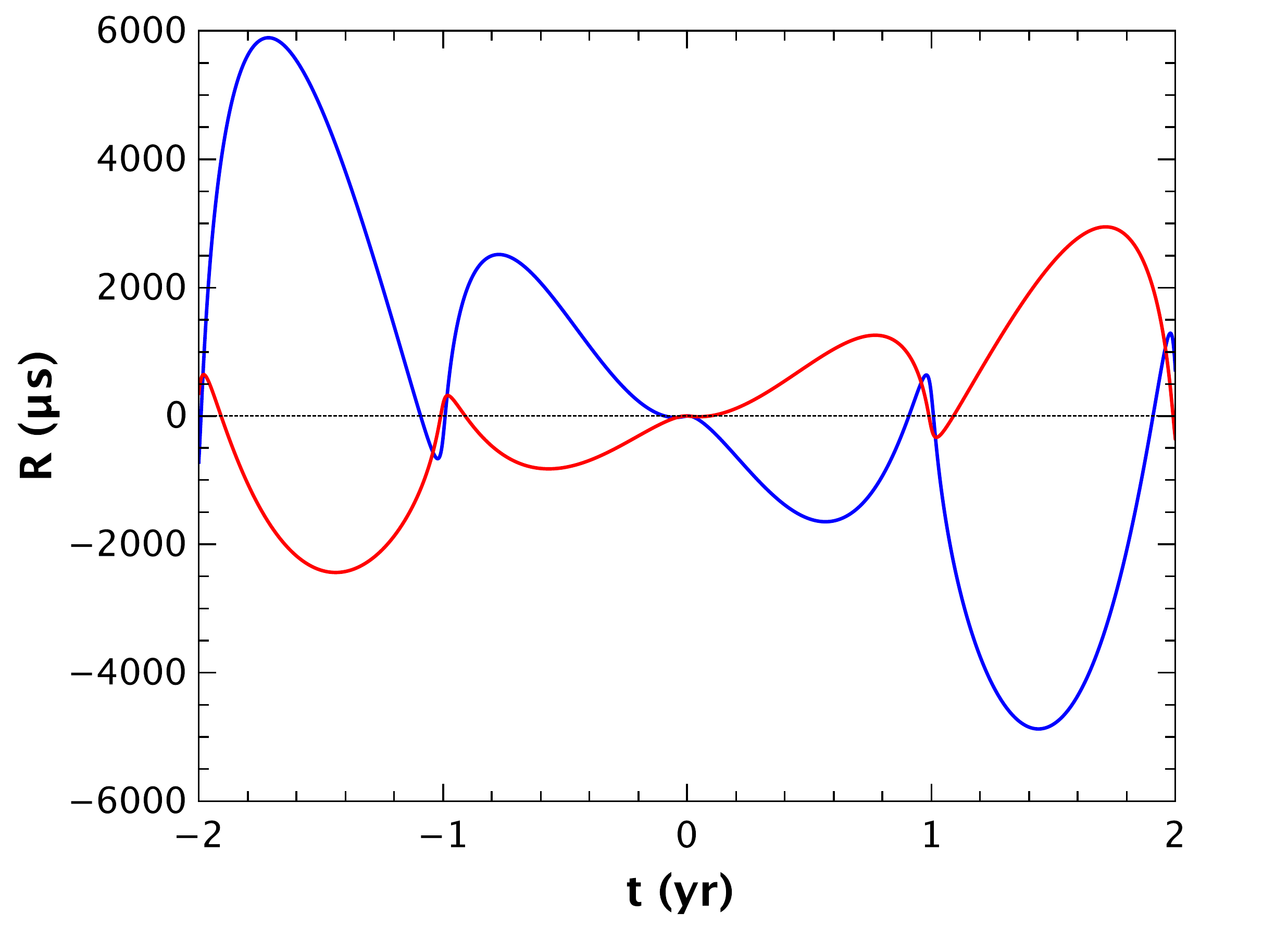}
\caption{Contribution to the Roemer delay caused by the proper motion of \sgr\
with respect to the SSB, for a 1\,yr orbit with an eccentricity $e =
0.8$. For the angles we have chosen $i = 60^\circ$, $\omega =
45^\circ$, and $\Omega = 0^\circ$ (blue) and $90^\circ$ (red).
\label{fig:DRpm}}
\end{center}
\end{figure}

Further constraints on the spin orientation, in particular on the
direction of the projection of the spin into the plane of the sky, do
come from the proper motion of \sgr\ with respect to the solar system 
barycenter (SSB). The transverse motion of \sgr\ with respect to
the SSB modifies the observed Roemer delay by
\begin{equation} \label{eq:DRpm}
  \Delta_{\rm R}^{\rm pm} = 
    \frac{1}{c} (\bldm{\mu}^\ast \cdot {\bf r})(t -t_0)\;,
\end{equation}
where $\bldm{\mu}^\ast$ is the angular proper motion vector of \sgr\
in the sky \citep{kop96}. The contribution $\Delta_{\rm R}^{\rm pm}$
has an impact on the arrival times of the pulsar signals which is
distinctly different from Lense-Thirring contributions (see
Fig.~\ref{fig:DRpm}), and depends on the orientation of the pulsar
orbit with respect to the well known proper motion
of \sgr\ \citep{rb04}. This can be easily demonstrated by looking at
the orbital averaged changes to the semi-major axis ($x$) and the
longitude of periapsis ($\tilde\omega$), which are given by
\begin{eqnarray}
  \dot x/x   &=&  \mu^\ast \cot i \sin\Omega \;,\\
  \dot{\tilde\omega} &=& -\mu^\ast \csc i \cos\Omega \;,
\end{eqnarray}
where $\Omega$ denotes the longitude of the ascending node (measured
clockwise in the sky, with respect to the direction of proper
$\bldm{\mu}^\ast$). Hence, the proper motion contribution to the Roemer
delay gives access to the sixth Keplerian parameter, i.e.\ $\Omega$,
and therefore completely determines the 3D orientation of the
orbit. Consequently, since we know the orientation of the black hole
spin with respect to the pulsar orbit from timing its orbital motion,
the 3D orientation of the black hole spin can be determined. This is
valuable input for combining pulsar observations with the measurements
of the \sgr\ shadow with the EHT, as we will discuss in
\S~\ref{sec:ehtpsr}. 

\subsection{Extracting the Quadrupole Moment From the Timing Residuals}
\label{subsec:psr.q}

Once the mass and spin are measured, a Kerr spacetime is fully
determined. Consequently, {as discussed above}, the measurement of any higher multipole moment is a test of the Kerr hypothesis. For a pulsar in
orbit around \sgr\ one can hope for the measurement of the quadrupole
moment, as the leading multipole moment, after $M_\bullet$ and
$S_\bullet$. The quadrupole moment of \sgr\ leads to a distinct signal
in the arrival times of the pulses, as it modifies the orbital motion
of the pulsar in a characteristic way \citep{wk99}. Based on
self-consistent mock data simulations, \cite{lwk+12} showed that, for
a pulsar with an orbital period of a few months, it should be possible
to extract the quadrupole of the \sgr\ spacetime from the timing
residuals with high precision. Depending on the rotation of \sgr\ and
the eccentricity of the orbit, this could be easily achieved with a
precision of $1\%$ or even better.

\begin{figure}[t]
\begin{center}
\includegraphics[width=8.5cm]{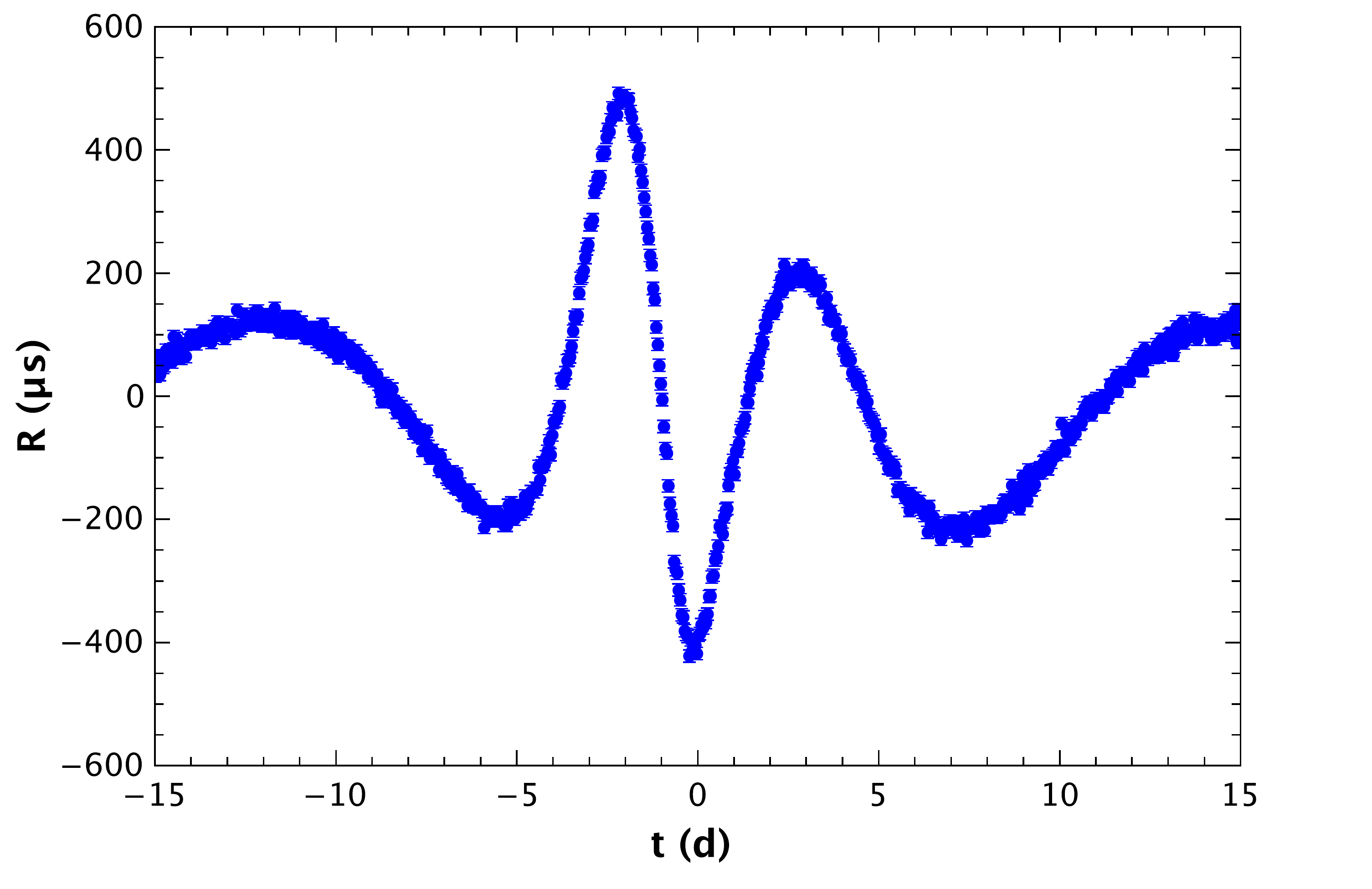}
\caption{Signature of the black hole quadrupole moment, for an
extreme Kerr black hole ($\chi = 1$). Simulations have been done for
two periapsis passages, with the above figure zooming into the first
one. 10\,$\mu$s TOAs have been created only within $\pm 15$ days
around the periapsis passages for a pulsar in an eccentric ($e = 0.8$)
orbit with $P_b = 0.5$\,yr. The orientation of the spin is
taken as in Figure~\ref{fig:res_so_3yr_09}. The residuals are a
result of a fit for the orbital and frame-dragging contributions. 
{For demonstration purposes we have used a high timing cadence, to 
densely map the quadrupole signature. In practice, such a coverage would be the result of many periapsis-passage observations over a few years.}   
\label{fig:Qsig1.0}}
\end{center}
\end{figure}

\begin{figure}[t]
\begin{center}
\includegraphics[width=8.5cm]{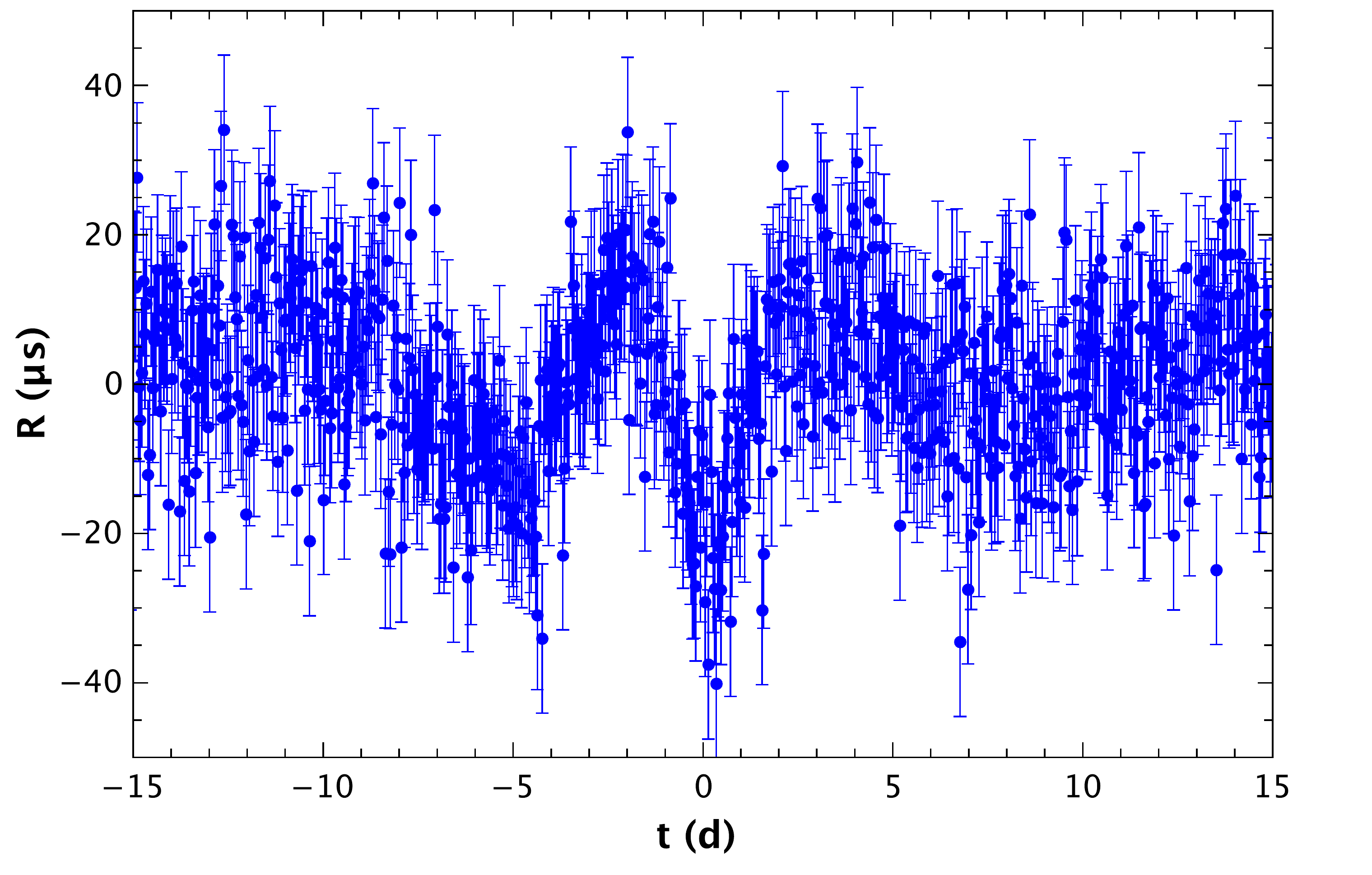}
\caption{Same as in Figure~\ref{fig:Qsig1.0}, but this time with $\chi =
0.2$. In this case, the quadrupole moment of the black hole leads to a
considerably less prominent signal in the residuals, but can still be measured
accurately given the assumed TOA error of 10\,$\mu$s.
\label{fig:Qsig0.2}}
\end{center}
\end{figure}

{The simulations for the no-hair-theorem test in \cite{lwk+12}
are based on the optimistic (cf.\ discussion
in \S~\ref{sec:gen}) assumption that the pulsar orbit does not
experience any relevant external perturbations and therefore the
secular precession of the orbit can be used to determine the spin of
\sgr.\footnote{Liu et al.~2012 have demonstrated a way to identify the
presence of external perturbations in the secular changes of the pulsar orbit.}
In this section we relax this assumption, like we have done above for the spin measurement. }
For our simulations we added the implementation for the
quadrupole moment of \cite{lwk+12} to our aforementioned extension of
TEMPO. Figure~\ref{fig:Qsig1.0} is the result of a simulation, where
we use timing data only near periapsis, and allow for an undetermined
overall precession of the orbit due to some unknown external
perturbations. Figure~\ref{fig:Qsig1.0} clearly shows that, after
fitting for the pulsar spin parameters, orbital parameters, and frame
dragging, there is still a distinctive signal in the residuals as a
result of the quadrupole moment of the black hole. As a general
result, depending on the timing-precision and the periapsis distance,
we will still be able to extract the quadrupole moment of \sgr. Of
course, this also depends on the actual value of the spin of \sgr, 
which is poorly constraint to date. In fact, the strength of the 
quadrupole effect scales with
$\chi^2$, and is therefore clearly less prominent for a slowly
rotating black hole (see Figure~\ref{fig:Qsig0.2}). Depending on the
timing precision, however, the quadrupole moment can still be
determined with high precision.

\begin{figure}[t]
\begin{center}
\includegraphics[width=8.5cm]{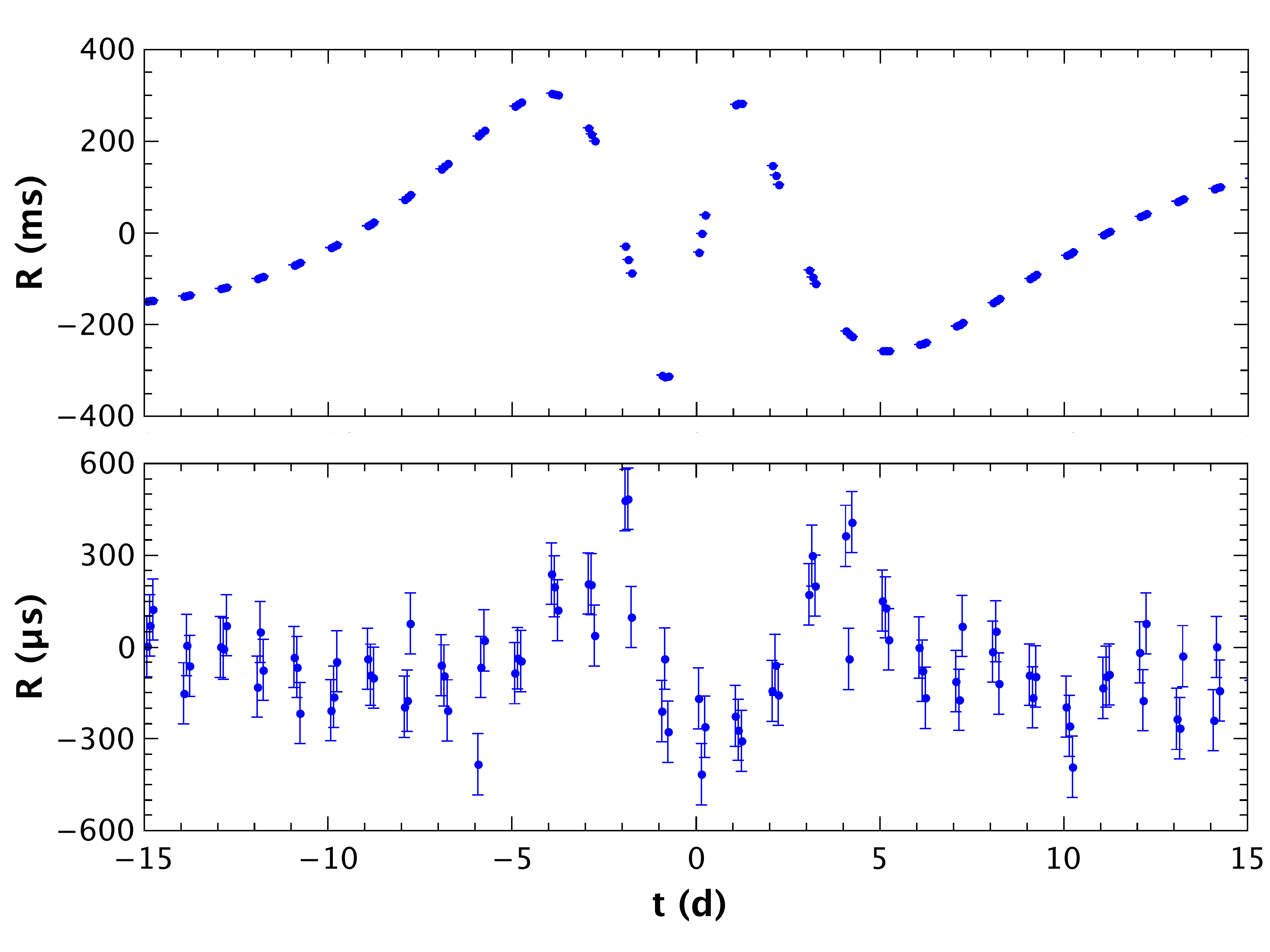}
\caption{Mock-data TOA ($\sigma_{\rm TOA} = 100\,\mu$s) coverage of 
the spin (top) and quadrupole (bottom) signal during a periapsis passage. 
The simulated data cover
three orbits.  We used $\chi = 0.6$, $P = 0.5$\,yr, $e = 0.8$, $\Theta
= 60^\circ$, $\Upsilon_0 = \omega_0 = 45^\circ$, and $\lambda =
55^\circ$.
\label{fig:res-S-Q}}
\end{center}
\end{figure}

\begin{figure}[t]
\begin{center}
\includegraphics[width=8.5cm]{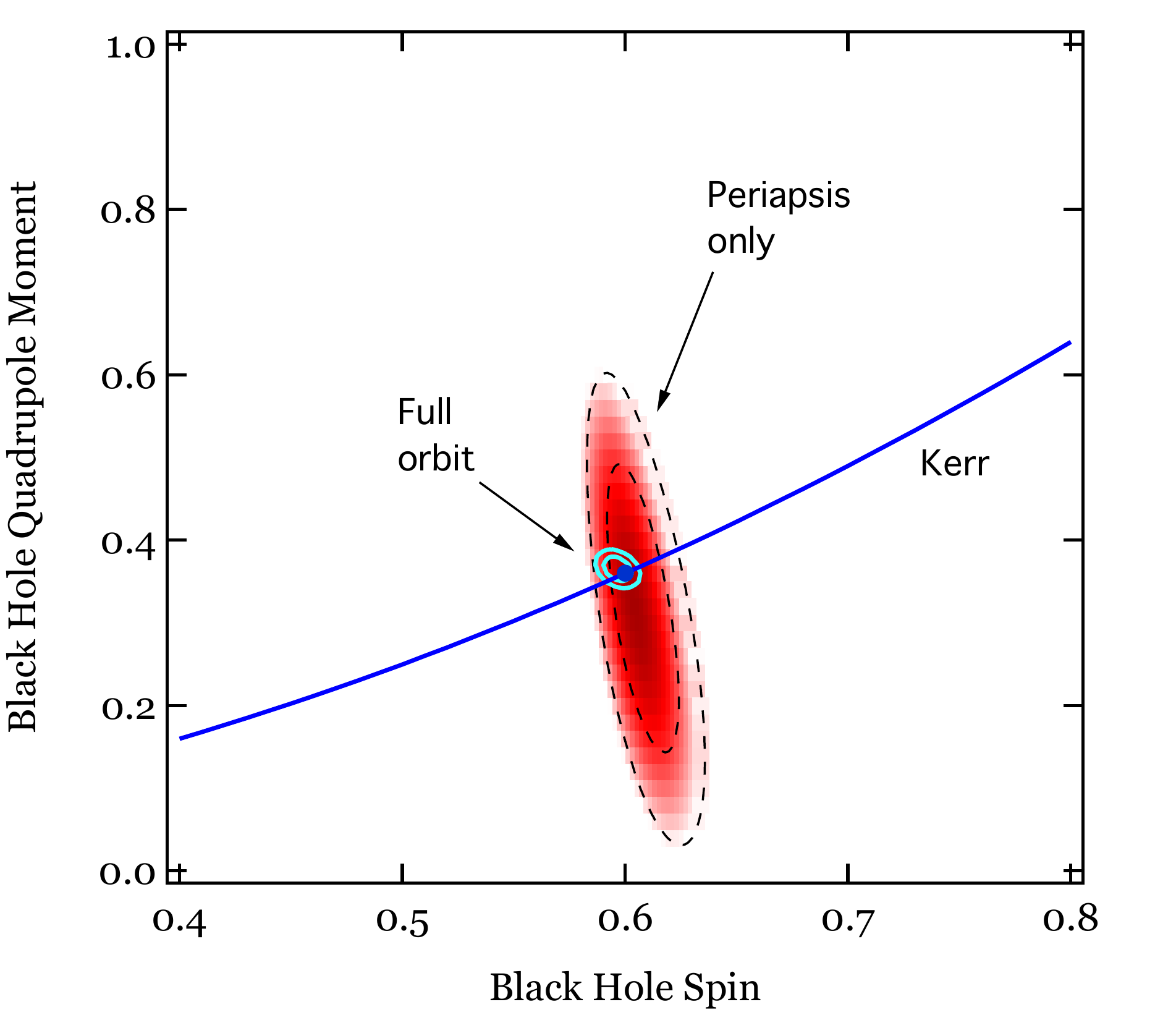}        
\includegraphics[width=8.5cm]{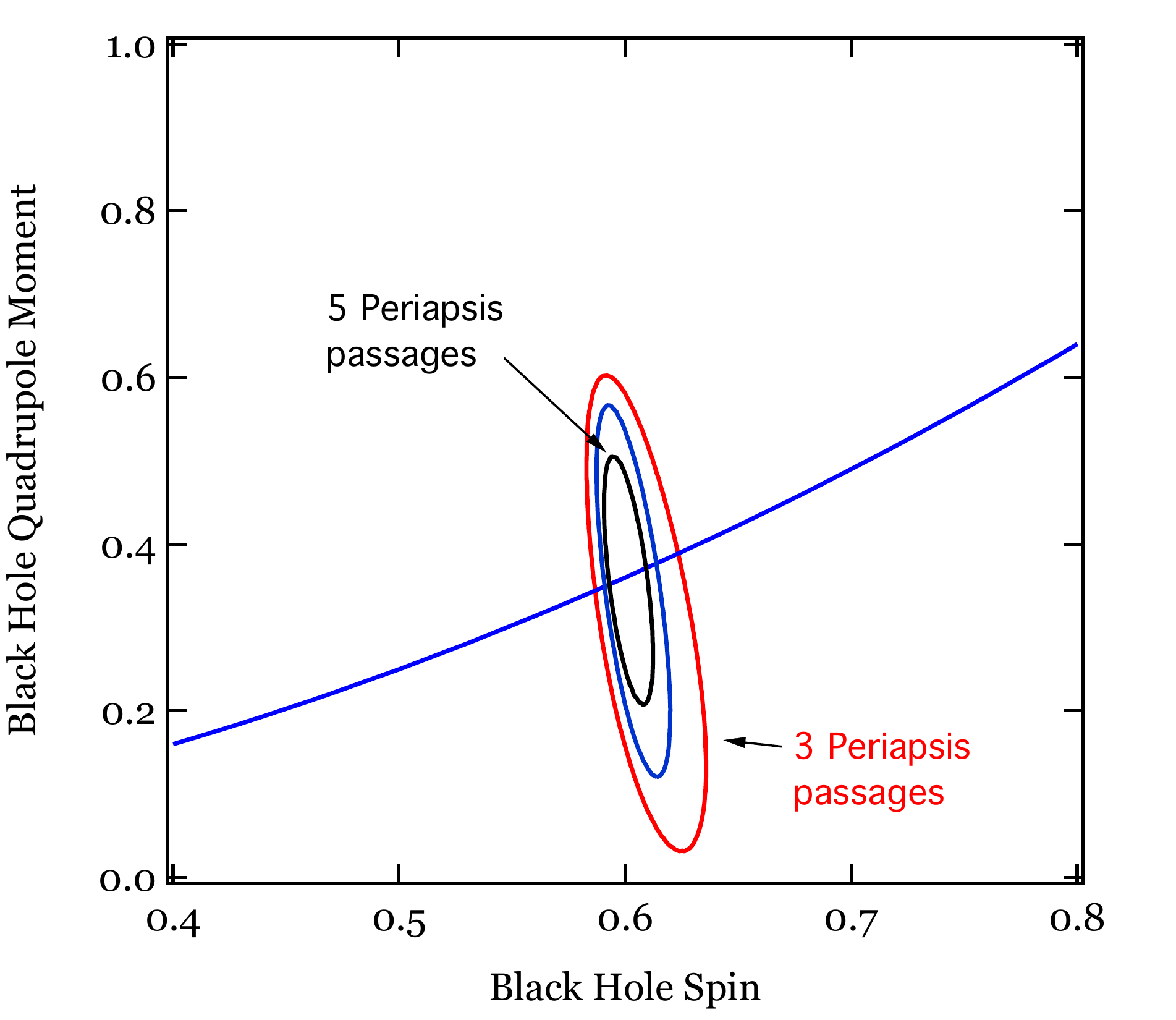}
\caption{The posterior likelihood of measuring the spin and quadrupole moment
of SgrA* using pulsar timing. In the top panel the dashed curves show
the 68\% and 95\% confidence limits while, in the bottom panel, the
solid curves show the 95\% confidence limits. The solid curve shows
the expected relation between these two quantities in the Kerr
metric. The filled circle marks the assumed spin and quadrupole moment
($\chi$= 0.6, $\vert q\vert$ = 0.36). The pulsar is assumed to have an
orbital period of 0.5\,yr (orbital separation of 2400\,$GM/c^2$) and
an eccentricity of 0.8, {while three TOAs per day with equal 
timing uncertainty of $100\,\mu$s have been simulated}. The top panel 
compares the uncertainties in the
measurement when only three periastron passages have been considered
in the timing solution to those when the three full orbits are taken
into account. The bottom panel shows the increase in the precision of
the measurement when the number of periastron passages is increased
from three to five.
\label{fig:psr_nohair}}
\end{center}
\end{figure}

We have conducted extensive mock data simulations to study the joint
measurability of spin and quadrupole moment. {Like in the simulations 
for the spin measurement, we have assumed three TOAs per day. 
Figure~\ref{fig:res-S-Q} shows the timing coverage of the spin and quadrupole 
signature during one periapsis passage}. Some of the results are
illustrated in the contour plots of Figure~\ref{fig:psr_nohair}. We
conclude that, {even for the conservative situation of a comparably 
low timing precision ($\sigma_{\rm TOA} = 100\,\mu$s) and the presence of 
external perturbations}, a quantitative test of the Kerr hypothesis is 
possible after only a few periapsis passages. If we have a better timing
precision or can make use of timing measurements along the whole
orbit, the spin and quadrupole moment can be determined with high
precision after a few orbits. The latter agrees with the findings
in \cite{lwk+12}.

\subsection{Distance Measurement with Pulsar Timing}

Given the large size of the pulsar orbit ($\sim 10^2\,{\rm au}$), the
orbital parallax \citep{kop95}, which is of order $\sim a^2/2cD$,
will lead to a significant contribution to the timing observations,
even for a moderate timing precision. This timing effect depends only
on well determined orbital parameters and the distance to \sgr, $D$,
and consequently can give independent access to $D$ (cf.~discussion
in Subsection~\ref{subsec:psr.mass}). The orbital parallax is a
periodic signal in the timing residuals, and therefore, if we have $N$
equally distributed TOAs with uncertainty $\sigma_{\rm TOA}$, its
measurement scales proportional to $\sigma_{\rm TOA}$ and
$\sqrt{N}$. Consequently we find
\begin{eqnarray}
  \delta D &\sim& 2\,\frac{c\sigma_{\rm TOA}}{\sqrt{N}}\,
                     \left(\frac{D}{a}\right)^2
  \nonumber\\
             &\sim& 20\,{\rm pc} 
                  \left(\frac{\sigma_{\rm TOA}}{10^2\,\mu{\rm s}}\right) 
                  \left(\frac{N}{10^3}\right)^{-1/2} 
                  \left(\frac{a}{10^2\,{\rm au}}\right)^{-2} \;, 
\end{eqnarray}
where we have used $D = 8.3$\,kpc.

External perturbations can also lead to changes of the orbit, which
could in principle partly mimic the above effects. This, however, depends
highly on the specifics of the perturbation, and we will not discuss
this in further detail in this paper. On the other hand, as argued
by \cite{lwk+12}, a precise measurement of the \sgr\ mass from pulsar
timing can be converted into a precise determination of the distance
to \sgr, when combined with high-precision astrometric observations in
the infrared. For instance, a high precision measurement of
$M_\bullet$ in combination with the (angular) size of the S2-star
orbit in \cite{2009ApJ...707L.114G} can be converted into a direct
measurement of the Galactic center distance with an error of
$\sim$100\,pc. Future 10\,$\mu$as astrometry promises a precision of
order one parsec or even better.


\section{Probing the spacetime of \sgr\ with the EHT}
\label{sec:eht}



The EHT will image the millimeter emission from \sgr\ with
horizon-scale resolution. There have been at least three proposals for
using EHT observations to map the spacetime of this black hole and, in
particular, to measure different combinations of its spin and
quadrupole moment.

The first approach utilizes the detailed shape of the shadow cast by
the black hole on the surrounding emission (Johannsen \& Psaltis
2010b). Because of the combined effects of frame dragging and of the
quadrupole deformation of the spacetime, the shadows of Kerr black
holes are nearly circular, independent of the black-hole spin and the
orientation of the observer (Bardeen 1973). The shadows of spacetimes
that violate the no-hair theorem, however, can be significantly
asymmetric, with the magnitude of asymmetry providing a measure of the
degree of violation of relation~(\ref{eq:chiqrel}); see Johannsen \&
Psaltis 2010b\footnote{A number of studies have explored the shapes
and sizes of black-hole shadows in modified gravity theories as well
as in parametrically modified Kerr-like metrics (see, e.g., Bambi \&
Freese 2009; Bambi \& Yoshida 2010; Johannsen 2012; Abdujabbarov et
al.\ 2013; Amarilla \& Eiroa 2013; Ghasemi-Nodehi et al.\ 2015).  In
this article, we focus on work that aims specifically to measure the
quadrupole moment of the black-hole spacetime using its shadow
properties.}.  The shape of the shadow can be measured using the
interferometric data either via an edge detection
scheme \citep{pnf+15} or via fitting phenomenological geometric models
(Ricarte \& Dexter 2015).

In a second approach, simulated images of the accretion flow are
fitted against the measured complex interferometric visibilities. The
characteristic scale of the brightness in the accretion flow is set by
the radius of the innermost stable circular orbit (see, e.g.,
Broderick et al.\ 2009; Dexter et al.\ 2010). For a general spacetime,
this radius is determined, in turn, by a particular combination of the
black-hole spin and quadrupole moment (see, e.g., Johannsen \& Psaltis
2010a). Even the current, limited imaging data at 1.3\,mm provide a
glimpse of how this method can be used to constrain the properties of
the black-hole spacetime by measuring the location of its innermost
stable circular orbit (Broderick et al.\ 2014) and this technique will
flourish as the full EHT array becomes operational.

{Finally, if either GRAVITY or the EHT finds evidence for
short-lived, compact emission regions (``hot spots'') that are
advected with the accretion flow, tracing their orbits will lead to a
measurement of the spacetime properties in a way that is similar to
those discussed in \S3 and \S4 for orbits of stars and pulsars
(Broderick \& Loeb 2006; Vincent et al.\ 2011). The dynamical
timescale in the vicinity of the horizon of \sgr\ is equal to a few
tens of minutes, i.e., much smaller than the time it will take to
generate an image. As a result, tracing the orbits of such hot spots
will be done by studying the time evolution of interferometric phases
or closure phases along appropriate baselines and baseline triangles
(Doeleman et al.\ 2009; Vincent et al.\ 2011).}

All three approaches have the potential of measuring different
combinations of the spin and quadrupole of the black hole.  However,
the first approach that involves measuring the shape of its shadow is
purely gravitational and, as such, is the least model-dependent: it
does not require a prior model of the accretion flow (as does the
second technique) and does not rely on assumptions about the advection
of particular compact emission regions along geodesics (as does the
last technique). For this reason, we will focus here on the first
approach as a proof of principle of our prospect of measuring the
properties of the black-hole spacetime with EHT observations.

Johannsen \& Psaltis (2010b) explored the asymmetry of the shadows of
spacetimes with independent spins and quadrupole moments. They used
the Glampedakis \& Babak (2006) spacetime, which is a formal solution
to the Einstein field equations up to the quadrupole order and remains
regular only for relatively slow spins (see discussion in Johannsen
2013). They devised an approximate relation that connects the
asymmetry of the shadow to the spin of the black hole, its quadrupole
moment, and the inclination of the observer, i.e.,
\begin{equation}
  A(\chi,q) = \Big[0.84\left(\chi^2+q\right) + 0.36\,\chi^3\Big]
              \left(\frac{GM_\bullet}{c^2}\right)\sin^{3/2}\lambda\;.
  \label{eq:EHT_asym}
\end{equation}
As this relation shows, when the no-hair theorem is satisfied, the
asymmetry depends only the third power of the spin, with a small
coefficient, becoming negligible for all but the fastest spinning
black holes. Measuring any appreciable asymmetry of the black-hole
shadow will be a strong evidence for a violation of the no-hair theorem.

\begin{figure}[t]
\begin{center}
\includegraphics[height=7cm,width=8cm]{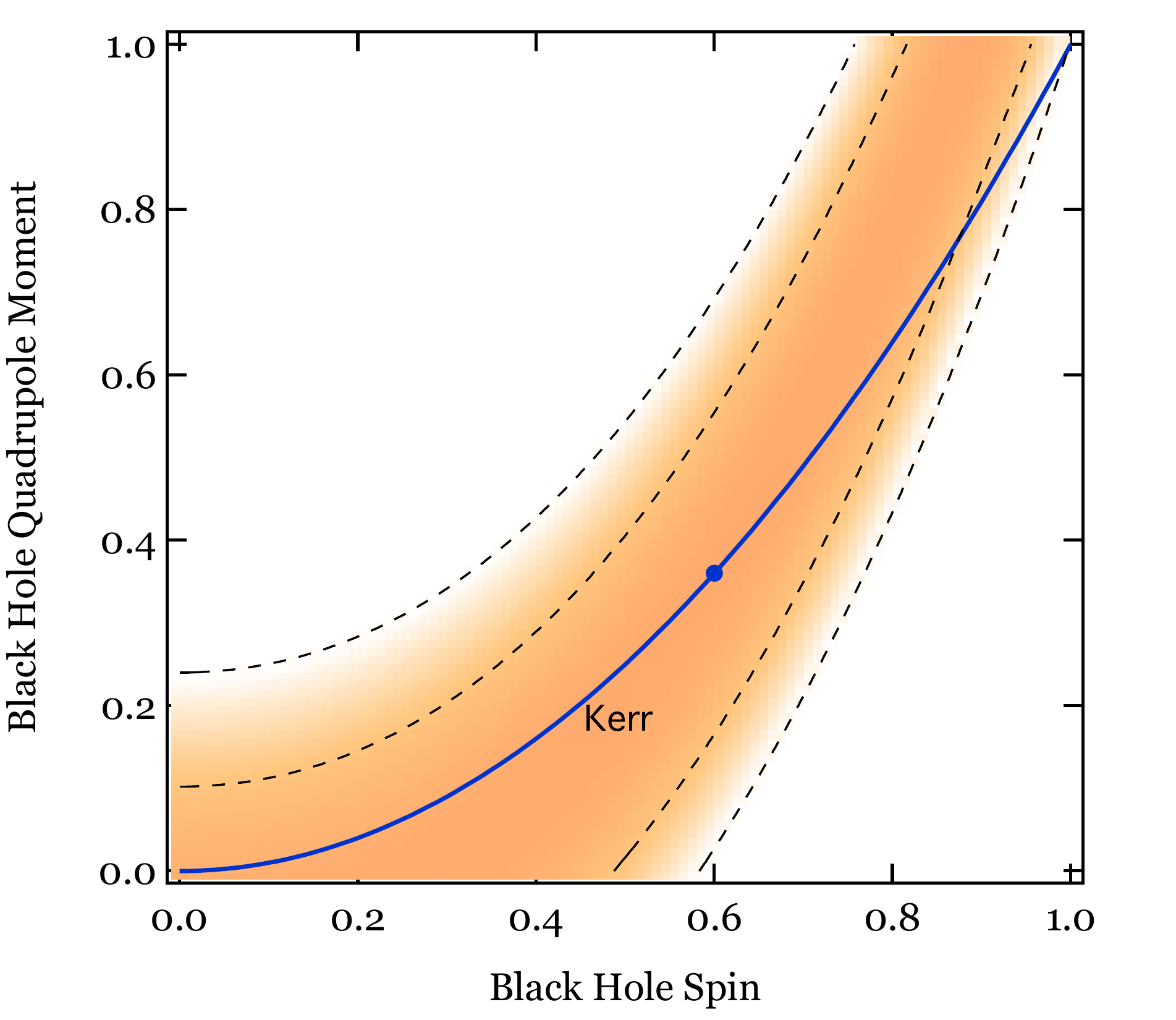}
\caption{The posterior likelihood of measuring the spin and quadrupole
  moment of \sgr\ using EHT observations of the shape of its
  shadow. The dashed curves show the 68\% and 95\% confidence
  contours, while the solid curve shows the expected relation between
  these two quantities for the Kerr metric.  The filled circle marks
  the assumed spin and quadrupole moment ($\chi=0.6$, $\vert
  q\vert=0.36$). As expected, the contours of maximum likelihood
  closely follow the Kerr relation, because any violation of the no
  hair theorem would have caused a measurable asymmetry in the shadow
  shape.\label{fig:eht_nohair}}
\end{center}
\end{figure}

The accuracy with which the shape of the black-hole shadow can be
measured with EHT observations will depend on the particular
techniques that will be used for image reconstruction and for pattern
matching. Johannsen et al.\ (2012) used approximate
relations for the flux of the photon ring that surrounds the
black-hole shadow as well as an extrapolation of the demonstrated
signal-to-noise ratio of existing EHT observations to infer that the
radius of the shadow can be measured to an accuracy of
\begin{equation}
  \sigma_{\rm rad} \simeq 4.3 \left(\frac{1\,{\rm mm}}{\lambda_{\rm obs}}\right)^2
  \left[\frac{53}{21}\left(\frac{1\,{\rm mm}}{\lambda_{\rm obs}}\right)-1\right]^{-1}
  \,\mu{\rm as} \;,
\end{equation}
where $\lambda_{\rm obs}$ is the observation wavelength.  In the full
EHT array, there are of order $\sim 9$ baselines with $u-v$
separations that are comparable to the position of the null due to the
shadow and with locations that are nearly uniformly distributed around
its circumference (see, e.g., Figure~3 of Ricarte \& Dexter
2014). These separations will allow us to measure the overall
asymmetry of the shadow along two axis with an accuracy of
$\sigma_{\rm A}\sim \sigma_{\rm rad}/\sqrt{9}\simeq 0.9\,\mu$as, where
we evaluated this last expression at a wavelength of 1.3\,mm. This
estimate is in agreement with the detailed study of Ricarte \& Dexter
(2014), who used mock EHT observations to show that the quality of the
data will allow measuring the properties of asymmetric crescents to an
accuracy that is smaller than a $\mu$as.

In order to visualize the correlated uncertainties in measuring the
spin and quadrupole moment of \sgr\ using this technique, we will
assume that EHT observations in the near future will lead to a
measurement of the asymmetry of the shadow of a Kerr black hole with a
spin $\chi=0.6$, inclinded at $\lambda=55^\circ$ with respect to the
observer. We will also assume a Gaussian posterior likelihood for this
measurement, with a centroid given by equation~(\ref{eq:EHT_asym}) for
$q=-\chi^2$, i.e., $A_0=A(0.6,-0.36)$, and a dispersion equal to
$\sigma_{\rm A}=0.9\,\mu$as. Then, using Bayes' theorem, we can write
the posterior likelihood that a given combination of a spin and
quadrupole moment are consistent with the data as
\begin{equation}
  P(\chi,q|{\rm data}) = P({\rm data}|\chi,q)P(\chi)P(q)\;,
\end{equation}
where $P(\chi)$ and $P(q)$ are the priors over the spin and
quadrupole, which we assume to be uniform between zero and one, and
\begin{equation}
  P({\rm data}|\chi,q) = \frac{1}{\sqrt{2\pi \sigma_A^2}}
    \exp\left\{-\frac{\left[A(\chi,q)-A_0\right]}{2\sigma_A^2D^2}\right\}\;.
\end{equation}

Figure~\ref{fig:eht_nohair} shows the resulting likelihood in the
spin-quadrupole moment parameter space. As expected, the contours of
maximum likelihood trace closely the Kerr relation $q = -\chi^2$,
unless the black hole has a very high spin, because any violation of
the no-hair theorem would have caused a measurable asymmetry in the
shadow shape.

It is important to emphasize that we have only considered here, in an
approximate fashion, the uncertainties related to the effective
resolution of the EHT images.  The uncertainty in our prior knowledge
of the ratio $GM/cD^2$ does not enter this measurement, because what
we will be measuring is the fractional asymmetry of the shadow shape
with respect to its overall apparent angular size. However, measuring
the shape of the black-hole shadow at the $\simeq 1$\% level requires
a prior knowledge, at a comparable level, of the properties of the
scattering screen that blurs the image. Our ability to characterize
the scattering screen at longer wavelengths and extrapolate its
properties down to the 1.3\,mm wavelength of the EHT observations will
be the main limiting factor in performing a test of the no-hair
theorem with the EHT (see discussion in Fish et al.\
2014; Psaltis et al.\ 2015a\nocite{pnf+15}).


\section{Combining the EHT experiment with stellar orbits and pulsar timing tests}
\label{sec:ehtpsr}



In the previous sections, we discussed in detail the constraints on
the measurement of the black-hole mass, spin, and quadrupole moment
that will be achieved in the very near future with upcoming
observations of stars, of pulsars, and of the black-hole shadow in
\sgr. Even though each type of observation may lead by itself to a
measurement of the black-hole properties, combining all three of them
offers an unprecedented advantage for three reasons.

First, each of these measurements will probe the spacetime at very
different distances from the black hole: pulsars and stars
will probe hundreds to thousands of gravitational radii, and the Event
Horizon Telescope will probe the inner tens of gravitational radii. If
a significant amount of matter is hidden very close to the black hole
in the form of dark matter particles or stellar-mass black holes,
these three different probes will allow us not only to constrain the
profiles of the hidden mass distribution but also to understand the
biases it introduces to the measurement of the black-hole properties.

Second, each of the measurements uses a very different observational
technique (e.g., astrometric positions of stars, timing of radio
pulsars, sub-mm images of the accretion flow) and is, therefore,
subject to very different systematic uncertainties. Comparing the
results from the three types of observations will allow us to
identify the systematics inherent to each.

\begin{figure}[t]
\begin{center}
\includegraphics[height=7cm,width=8cm]{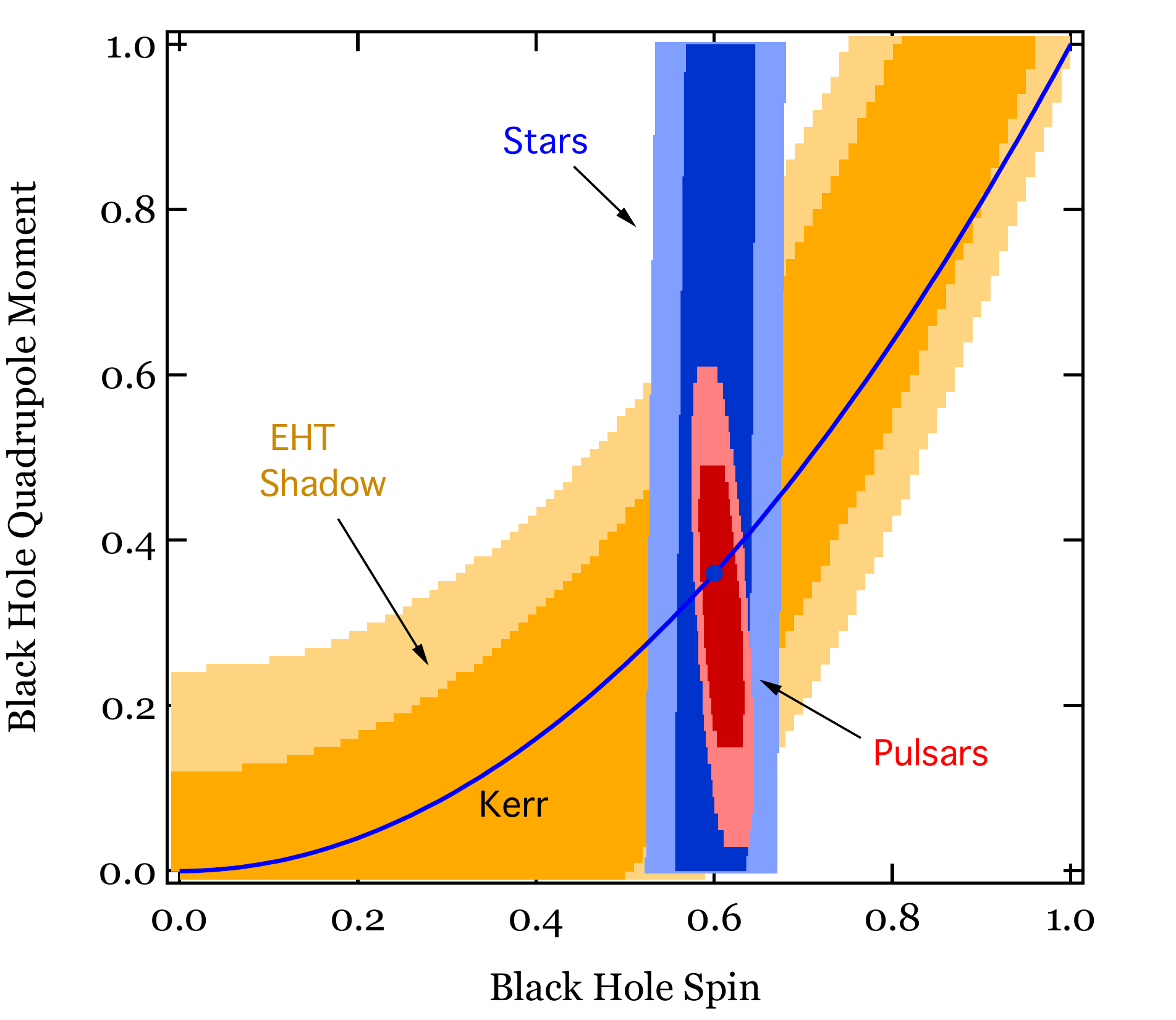}
\caption{Comparison of the posterior likelihood of measuring the spin
  and quadrupole moment of \sgr\ using the orbits of two
  stars {\em (blue)\/}, timing of three periapsis passages of
    a low-precision pulsar {\em (red)\/}, and the shape of its shadow
  {\em (gold)\/}. The solid curve shows the expected relation between
  these two quantities for the Kerr metric. The filled circle marks
  the assumed spin and quadrupole moment ($\chi=0.6$, $\vert
  q\vert=0.36$). Combining these three independent types of
  measurements, each of which suffers from different biases and
  potential systematic uncertainties, will significantly increase our
  confidence in the inference of these two black-hole properties and
  in the test of the no-hair theorem.\label{fig:all_nohair}}
\end{center}
\end{figure}

Finally, each type of observations is expected to lead to correlated
uncertainties (or even degeneracies) between the black-hole spin and
quadrupole moment. However, the correlations and degeneracies in each
method are along different directions in the parameter space (see
Figure~\ref{fig:all_nohair}). The orbital precession of stars and
pulsars will measure primarily the spin of the black hole. The timing
of pulsars will measure independently the quadrupole moment of the
spacetime. A detection of an asymmetry of the black-hole shadow will
measure deviations of the quadrupole moment from the Kerr value.
Combining all these measurement will lead to uncorrelated
measurements of the black-hole spin and quadrupole moment and hence
provide a test of the gravitational no-hair theorem.

Even though our focus in this article has been on testing the no-hair
theorem, it is also important to emphasize that combining these three
types of measurements will also have other important implications for
the astrophysics of accretion flows and of supermassive black holes in
the centers of galaxies. For example, as discussed above, differences
among the enclosed mass inferred at different radii with the EHT, with
stars, and with pulsars, will allow us to constrain the distribution
of stellar objects and dark matter at the very center of our
Galaxy. Furthermore, measurement of the relative orientation of the
black-hole spin and the angular momentum of the inner accretion flow
will inform our understanding of black-hole feeding and alignment of
black-hole spins (see, e.g., the discussion in Psaltis et al.\ 2015a).

It is true that the EHT and GRAVITY experiments still need to
demonstrate that they can operate at their designed specifications and
a pulsar, as well as at least two stars need to be discovered in
sufficiently close orbits around \sgr\ for the three types of
observations discussed here to be realized at the required
accuracy. However, all these are possible in the very near future and
promise to revolutionize our probes and understanding of strong-field
gravity and accretion flows in the vicinity of black holes.


\acknowledgements

DP thanks the Max-Planck-Institut f\"ur Radioastronomie for their
hospitality during the visit in which this project was conceived. DP
acknowledges support from NASA/NSF TCAN award NNX14AB48G and NSF grant
AST~131203. NW acknowledges valuable discussions with Tal Alexander
and Kuo Liu. This work has been supported by the ERC Synergy Grant
{\em BlackHoleCam} under grant agreement no.~610058. This research was
also supported by the Munich Institute for Astro- and Particle Physics
(MIAPP) of the DFG cluster of excellence ``Origin and Structure of the
Universe''.

\bibliographystyle{apj}

\end{document}